\documentclass[11pt,a4paper]{article}

\usepackage[utf8]{inputenc}
\usepackage[T1]{fontenc}
\usepackage[margin=1in]{geometry}
\usepackage{microtype}
\usepackage{amsmath,amssymb}
\usepackage{graphicx}
\usepackage{hyperref}
\usepackage{cite}
\usepackage{authblk}
\usepackage{abstract}
\usepackage{enumitem}
\usepackage{xcolor}
\usepackage{booktabs}
\usepackage{lineno}
\newif\ifdraft
\draftfalse   

\newcommand{\draftversion}{v1.0rc}
\providecommand{\draftrecipient}{Final draft for submission}   

\ifdraft
  \usepackage{lineno}
  \usepackage{fancyhdr}
  \usepackage{xcolor}
  \usepackage{draftwatermark}

  \SetWatermarkText{\draftversion}
  \SetWatermarkScale{1.9}
  \SetWatermarkColor[gray]{0.9}   

  \newcommand{\draftrecipientline}{%
    \ifx\draftrecipient\empty\else
      \space\textbar\space Prepared for: \draftrecipient
    \fi
  }

\fi
\usepackage{caption}
\usepackage{mdframed}
\newmdenv[
  linewidth=0.6pt,
  linecolor=black!40,
  leftline=true,
  rightline=false,
  topline=false,
  bottomline=false,
  innerleftmargin=8pt,
  innerrightmargin=0pt,
  innertopmargin=2pt,
  innerbottommargin=2pt,
  skipabove=4pt,
  skipbelow=4pt
]{regime}
\hypersetup{
    colorlinks=true,
    linkcolor=black,
    citecolor=black,
    urlcolor=blue!60!black
}

\usepackage{todonotes}

\title{From Newtonian to Relativistic IAM:\\
\large The Autonomous Principal as Reference Frame for Digital Identity}

\author[1]{Philippe Page}
\author[1]{Robert Mitwicki}
\author[2]{Michal Pietrus}
\affil[1]{\textit{Human Colossus Foundation}}
\affil[2]{\textit{Argon\textbf{auths}}}

\date{\today}

\begin{document}
\ifdraft\linenumbers\fi

\maketitle

\begin{abstract}
\noindent
The 2023 paper \emph{Distributed Governance: a Principal-Agent Approach to Data Governance} \cite{page2023distributed} introduced the autonomous principal as the locus of transactional sovereignty in digital ecosystems. This follow-up advances a structural argument for why that model is not a normative preference but a consequence of taking causality seriously in distributed information systems. Drawing an analogy with the transition from Newtonian to relativistic physics, we show that custodial identity management rests on an implicit assumption of global simultaneity that fails as soon as identity must operate across ecosystems, jurisdictions, and the offline/online boundary. Once that assumption is dropped, state ceases to be a noun held by a central authority and becomes a relation maintained between principals through causally ordered exchanges. The autonomous principal emerges as the only entity with standing to define its own reference frame. We report on technology built since 2023 that operationalises this view, and outline its consequences for cross-border data flows and agentic systems.
\end{abstract}
\tableofcontents
\newpage
\bigskip

\section{Introduction}
\label{sec:intro}

In \cite{page2023distributed} we proposed a distributed governance model built around \emph{autonomous principals}: entities capable of choice, and therefore capable of exercising transactional sovereignty within and across ecosystems. We defined ecosystems as units of governance, extended the legal notion of the privacy sphere into the digital domain, and outlined how a digital self can carry rights and accountability across jurisdictional boundaries. The argument there was principally drawn from principal-agent theory and legal philosophy.

This paper makes a different argument for the same conclusion. We claim that the autonomous-principal model is not merely a desirable governance design but the structurally inevitable consequence of how \emph{state} --- the information that defines who a principal is, what they may do, and what they have consented to at any given moment --- behaves in distributed information systems. The argument proceeds by analogy with the transition from Newtonian to relativistic physics: just as Maxwell's equations forced physics to abandon absolute simultaneity, multi-ecosystem digital interaction forces identity and access management (IAM) to abandon \emph{custodial state} --- state held by an authority on the principal's behalf, against which the principal's identity is checked at the moment of any transaction. What survives the collapse of the custodial assumption is precisely the autonomous principal of \cite{page2023distributed}, now understood as the reference frame from which a digital self is authoritatively defined.

The paper has two contributions. First, a structural argument (\S\ref{sec:argument}) that grounds the autonomous-principal model in distributed-systems theory rather than in normative governance preferences. Second, a report (\S\ref{sec:technology}) on infrastructure built since 2023 that operationalises this view, together with two worked applications (\S\ref{sec:applications}). We position the work against adjacent frameworks --- in particular self-sovereign identity (SSI) and blockchain-anchored identity --- in \S\ref{sec:not-this}.

\section{The structural argument}
\label{sec:argument}

\subsection{Newtonian IAM: the assumption of a global \emph{now}}
\label{sec:newtonian}
Today's identity infrastructure --- enterprise directory services, federated identity providers, national electronic identity schemes, and the issuer-holder-verifier model of self-sovereign identity \cite{w3c-did,w3c-vc,allen2016path} --- shares a common architectural assumption: that at any moment, the canonical state of a subject (who they are, what attributes they hold, what they have consented to) resides with some authority capable of being consulted. The authority may be a single enterprise, a federation, a state, or a credential issuer; the holder may or may not control disclosure of credentials issued against that state. In every case, however, a privileged vantage point exists from which the subject's state is authoritatively known. We call this the \emph{Newtonian assumption}: the implicit positing of an absolute frame against which identity facts can be ordered and resolved.
\begin{figure}[htbp]
    \centering
    \includegraphics[width=0.78\textwidth]{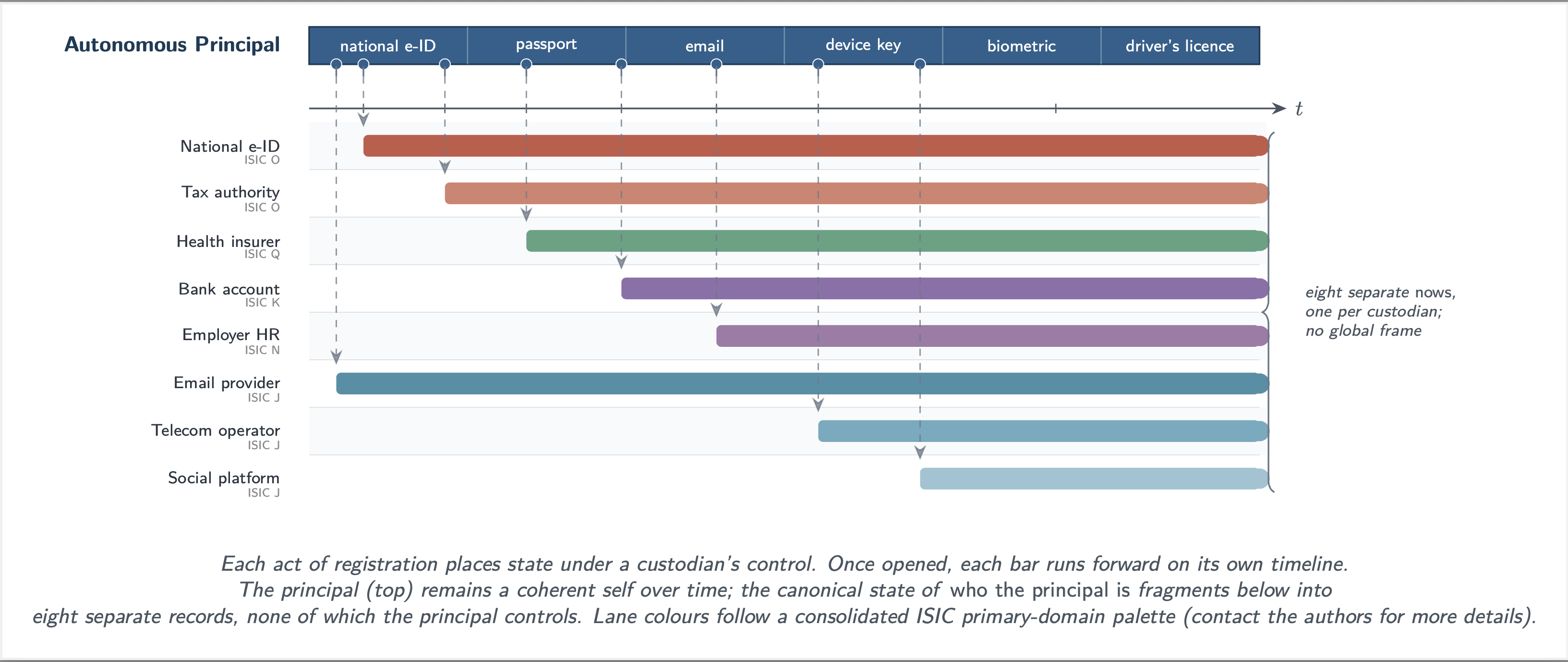}
    \caption{Custodial identity over time. The principal (top band) remains coherent; the canonical state of who the principal is fragments below into custodial records, each running on its own timeline; the dashed downward arrows indicating the start of the custodial state. Lane colours follow a consolidated ISIC primary-domain palette.}
    \label{fig:custodial-gantt}
\end{figure}

The choice of name carries the analogy's full weight, and we mean it generatively. Newtonian mechanics is not a discarded theory; it is the working physics of nearly every engineered system in daily use, from the aircraft we board to the bridges we cross. Relativity did not invalidate it. Relativity extended physics into regimes Newton could not have anticipated --- regimes whose practical exploitation gave us satellite positioning, semiconductor devices, and the materials science of the modern world. The relationship we propose between custodial and relational identity is of the same character. Custodial IAM is the working identity infrastructure of the digital economy; it runs the enterprise, the bank, the hospital, the state. We are not proposing its replacement. We are proposing an extension into regimes it was never designed to address.

What characterises those regimes is the absence of a privileged vantage point. Identity is now routinely required to operate across jurisdictions whose regulatory regimes conflict by design, across ecosystems whose authorities do not recognise one another, across the offline/online boundary, and --- with increasing urgency --- across human and agentic principals whose interactions admit no single administrative authority. In these regimes the question \emph{what is the canonical state of this subject?} no longer has a well-defined answer, not because any custodian has failed, but because the question presupposes a frame that the regime does not provide. The Newtonian assumption does not become wrong; it becomes inapplicable, in the same precise sense that Newtonian physics is the limit case of special relativity, applicable to phenomena involving relative motion at velocities far below $c$, the speed of light in vacuum.

Self-sovereign identity is best understood, in this light, as the opening move of the necessary extension. By relocating control of credential disclosure to the holder, SSI broke the assumption that the custodian must also mediate the subject's interface to the world --- a genuine and consequential advance. What the issuer-holder-verifier model retains is the deeper assumption that an authority somewhere holds the canonical state against which credentials are issued. The argument of this paper is that completing the extension requires going further: state itself, not merely its disclosure, must become relational. This is not a critique of the SSI programme but a continuation of its trajectory into the regime where its founding intuition --- that the subject is the proper locus of authority over their digital self --- reaches its full structural expression.

\subsection{Where the assumption breaks}
\label{sec:fails}

Having located the structural assumption in \S\ref{sec:newtonian}, we identify three regimes in which it ceases to be applicable. These are not vendor failures or implementation bugs; they are the boundaries of the domain of validity, in the same way that the failure of Newtonian addition of velocities at relativistic speeds is not a defect of Newton but a marker of his theory's reach.

\begin{regime}
\textbf{(a) The cross-jurisdiction regime.} Two or more legitimate authorities each hold canonical state for the same subject under conflicting governance (GDPR vs.\ sectoral US law; EU member-state implementations of the same directive; healthcare records mirrored between provider and national registry). There is no principled basis within the custodial model for choosing between them.
\end{regime}

\begin{regime}
\textbf{(b) The propagation-lag regime.} The custodian's record diverges from the operative truth because revocations, consents, and offline events propagate slowly or not at all. The record persists; the relation has moved on. Crucially, no amount of synchronisation engineering closes this gap, because some events --- particularly offline ones and those crossing administrative boundaries --- have no channel to the custodian by design.
\end{regime}

\begin{regime}
\textbf{(c) The constituting regime.} The relation between principal and authority is itself what needs to be governed, not presupposed --- guardianship being the canonical case (\S\ref{sec:guardianship}). The custodial model has no place to put a relation that constitutes a holder, because it assumes the holder.
\end{regime}

\begin{regime}
\textbf{(d) The agentic regime.} The principal acting in a relation may be an autonomous agent --- a software entity instantiated to act on behalf of a natural or legal person, often only for the duration of a single task. Such an agent has no pre-existing identity record for any custodian to consult, its authority is delegated and may evolve faster than any custodial record can track, and the principal on whose behalf it acts may itself be another agent, producing chains of delegation that custodial models have no clean way to represent. The 2024--2026 literature on agent governance is in substantial part an attempt to find the missing structural primitive that would let identity infrastructure carry these cases. We touch on the architecture's response in \S\ref{sec:guardianship} and outline a programme of further work in \S\ref{sec:conclusion}.
\end{regime}

Each of these is treated  in the applications of \S\ref{sec:applications}. For now, the structural point is sufficient: there exist regimes in which no single authority can hold canonical state, not as a contingent matter of engineering but as a structural matter of what state is in such regimes.

\subsection{State as relation: lessons from distributed systems}
\label{sec:relation}

The problem of ordering events in distributed systems is not new. In 1978 Leslie Lamport defined the \emph{happens-before} relation \cite{lamport1978time}, drawing explicitly on the causal structure of special relativity: in the absence of a global clock, the only well-defined order between events is the partial order induced by the causal pathways through which information can flow. The paper opened a research programme that has extended this insight to increasingly complex distributed regimes, and gave the field its most enduring conceptual import from physics.

Our claim is that the same insight, taken seriously, restructures digital identity. In the highly distributed and hyper-connected regimes diagnosed in \S\ref{sec:fails}, causal ordering is one of the elements essential to fidelity --- but fidelity is not yet identity. Identity adds four further constraints that custodial architectures struggle to meet. The first is that current digital identity sits at the edge of the distributed system: identifiers are bound to identity holders (the principals) through third-party custodians, and the principal loses control of the binding to the custodian. The second is that identifiers --- email addresses, account names, national registry numbers --- are difficult to update at the speed at which the underlying status changes, and impossible to update without the custodian's participation. The third, and structurally the most important, is that the relation between the principal and the authority is itself contextual. The constituting regime under which a relation comes into existence cannot be known in advance; it is determined by the relation itself. Identity cannot remain in the custody of a third party in such regimes, because no third party is in a position to define the relation on the principal's behalf.

The fourth, and now most acute, is that the principal acting in a relation may not be an autonomous principal at all but an agent acting on behalf of one --- a software entity whose state is not antecedent to the relation but constituted by the delegation chain that traces back to an autonomous principal in the sense of \cite[Defs.~1.1--1.3]{page2023distributed} (natural person, legal person, or governmental authority). The chain is the state. No custodian can hold canonical state for such an agent because the state is not a property of the agent but a property of the path of delegation, which crosses jurisdictions, ecosystems, and authority levels that no single custodian encompasses.

This paper argues, in consequence, that the principal must be the locus of its own reference frame --- the origin from which its digital identity is authoritatively defined. The subtitle of the paper, \emph{relativistic IAM}, is meant in the precise sense Lamport's debt to physics establishes: digital identity is relative to the interactions between an autonomous principal and the other entities of the system, and it is transactional in the sense that its state evolves through the causally ordered sequence of those interactions.

\paragraph{Applying the intrinsic/extrinsic decomposition to identity.} Part~1 \cite{page2023distributed} introduced a structural distinction between the intrinsic and extrinsic properties of data, drawing on the same kind of analogy with physics that the present argument has used elsewhere: the mass of an object is intrinsic, but its weight depends on the gravitational field it inhabits. In Part~1's framework, the intrinsic properties of a data object are defined by its data model and are independent of usage; the extrinsic properties are supplied by the governance and economic domain of the ecosystem in which the data object acquires meaning, value, or role. Their composition is summarised in Part~1's Equations~(6) and~(7): \emph{data} are intrinsic properties, and \emph{information} is data plus extrinsic properties. The decomposition is made operationally tractable by decentralised semantics (objectual integrity) and decentralised authentication (factual authenticity), which together let any party verify the intrinsic side without recourse to a custodian.

\paragraph{Why this decomposition matters beyond data.} The intrinsic/extrinsic distinction is a stronger conceptual primitive than the data/metadata distinction that dominates current vocabulary. Data/metadata depends on a fixed boundary --- the metadata describes the data --- that breaks down whenever the boundary itself is contextual: a field that is metadata in one transaction is data in another, a catalogue entry that is metadata to its records is data to the registry that catalogues catalogues entries. The intrinsic/extrinsic decomposition is, by contrast, defined relationally: properties are intrinsic to an object \emph{relative to a frame of reference} and extrinsic relative to the field in which that frame is embedded. The decomposition is fractal --- what is extrinsic at one level becomes intrinsic when the level shifts, since the governance frame of an ecosystem (extrinsic to a data object within it) is itself intrinsic to that ecosystem when the ecosystem is the unit of analysis. This generativity is what lets the same primitive carry the argument across the data layer (Part~1) and into the identity layer (the present paper), and we expect it to extend further as the relational architecture is applied to other domains.

The same decomposition applies, without modification, to identity. An \emph{identifier} carries the intrinsic properties of a principal's digital existence: the cryptographic provenance of its control material, the content-addressed schema under which it is recognised, the causal log of events through which its control has evolved. These properties are stated here as requirements; the protocols that meet them are described in \S\ref{sec:technology}. They are independent of any particular ecosystem in which the principal participates; they are properties of the principal as such, verifiable by any party in possession of the relevant artifacts. An \emph{identity}, on the other hand, is information in Part~1's precise sense: the identifier together with the extrinsic properties that a particular ecosystem's governance frame assigns to it --- rights, duties, roles, recognised attributes, the operational standing the principal has within that ecosystem. The same identifier can acquire different identities across different ecosystems, not because the principal has changed but because the extrinsic frame has.

This is the structural reason no custodian can hold canonical identity-state across ecosystems. The custodian could, in principle, hold the intrinsic side; but the intrinsic side does not need a custodian, because it is end-verifiable by construction. The extrinsic side cannot be held custodially, because it is by definition supplied by the ecosystem's governance frame, which is plural across the regimes of \S\ref{sec:fails} and irreducible to any single authority's record. Identity, taken as information, is what emerges from the interaction of an intrinsically recognisable principal with the extrinsic frame of a particular relation. It is not held; it is composed, and it is composed differently in each relation.

\paragraph{The operational consequences.} The operational implications for a system designed on this basis follow from the three requirements that any relational identity architecture must meet: integrity (the data defining the principal's intrinsic side must be cryptographically verifiable without recourse to a custodian), authenticity (control over an identifier must be demonstrable end-to-end, with the chain of control evolution preserved), and governance (the rules under which relations are constituted and evolve must themselves be artifacts of the relation, not external assumptions). The protocols developed in this work to meet these three requirements jointly are described in \S\ref{sec:technology}; we note here only that the distributed-systems literature has accumulated substantial relevant material, much of which addresses one or another aspect of the trilogy. Conflict-free replicated data types \cite{shapiro2011crdts} address convergence of replicated state without coordination, an integrity-adjacent concern; consensus protocols from Paxos onward \cite{lamport1998paxos} address the cost of manufacturing global ordering where it is genuinely required, a governance-adjacent concern; append-only causal logs underpin event-sourced architectures and provide the temporal substrate over which authenticity claims can be made. None of these primitives addresses the trilogy in full --- the integrative work is what the architecture of \S\ref{sec:technology} undertakes.

The general lesson, common to relativity and to distributed systems, is the same: once instantaneous global information is forbidden, state stops being a noun held somewhere and becomes a relation maintained between participants through causally ordered exchanges. The structural argument of this paper is that digital identity has entered the regime in which this lesson applies, and that the architecture of identity must follow.

\subsection{The autonomous principal as reference frame}
\label{sec:frame}

The structural argument can now be stated in its strongest form. \emph{The autonomous principal is not a participant in IAM; it is the reference frame from which IAM becomes coherent.} The three preceding subsections have prepared the ground for this claim; what remains is to draw out its consequences.

\paragraph{No external \emph{now}.} The hidden assumption beneath custodial identity, never named in those terms by its practitioners, is that an external \emph{now} exists --- a privileged moment, sustained by some custodian, from which the principal's state can be authoritatively defined. The first three regimes of \S\ref{sec:fails} show this assumption breaking under sustained pressure: cross-jurisdiction (multiple nows under conflicting governance, no principled basis to choose between them), propagation-lag (the custodian's now diverges from operative truth), and the constituting regime (the relation precedes any canonical now from which to define it). The custodial-state figure (Fig.~\ref{fig:custodial-gantt}) made this visible: each custodian carries its own now, none shared, each connected to the principal under bindings of widely varying strength --- from cryptographically grounded identifiers issued by recognised authorities down to nominal identifiers anyone can self-assert --- and none of those bindings under the principal's own control or systematically reconcilable across custodians. The agentic regime sharpens the impossibility further --- the three failure modes combine at machine speed in a delegation chain whose every link demands a now that cannot be supplied by any party other than the principal at the head of the chain. Once it is seen that no external now can be sustained across these regimes, only one structural conclusion remains: the principal must hold the frame from which its own state is defined.

The temporal scope of the relevant \emph{now} ranges from the quasi-instantaneous to the lifetime, and ranges over every kind of autonomous principal Part~1 defined \cite{page2023distributed}. For a natural person, the scope of a single access token may be bounded in seconds, while the scope of a birth certificate is bounded by the principal's own existence. For a legal person such as a corporation, the scope of a single client interaction may be measured in milliseconds, while the scope of the corporate person itself extends from registration to dissolution --- a period that may span decades and traverse mergers, restructurings, and jurisdictional moves. For a governmental authority, the corresponding scope reaches from the transaction to the constitutional lifetime of the institution. The structural point holds across this entire range and across all classes of principal: whatever the scope, the locus from which the state is defined is the relation itself, not a custodian standing outside it. A custodian asked to define a principal's state at any moment is, in effect, claiming a privileged temporal frame; the relativistic argument of \S\ref{sec:relation} is that no such frame exists.

\paragraph{The principal must hold the frame.} It follows that the principal must hold the reference frame from which its own state is authoritatively defined. Only the principal can meaningfully navigate the multiple ecosystems in which it participates. Consider a Spanish national, born in Australia, holding a Swiss driving licence, renting a car in the United States: four jurisdictions, four governance frames, four sets of extrinsic properties supplied by ecosystems whose authorities do not stand in any hierarchical relation to one another. The example concerns an individual, but individuals are only one kind of autonomous principal in the sense of Part~1, and the multi-jurisdictional case is not exceptional --- it is constitutive of entire domains. Global trade and supply-chain ecosystems operate under multiple regulatory and certification authorities by design, and the coherent identity of a single shipment, batch, or product across those authorities is everyday operational reality. Health-related information is governed across jurisdictional and institutional boundaries whose plurality is recognised in policy --- the European Health Data Space (EHDS) being the most visible current expression --- and structural in practice; we treat one operational instance, described in its own context as a ``mini-EHDS,'' in \S\ref{sec:nextgen}.  In these domains the question is not whether the principal will encounter plural governance frames but how the architecture handles a plurality that is already there. No custodian holds the canonical state of the principal across the different frames, because no such custodian could; the relation between the principal and each ecosystem is what defines the principal's standing within that ecosystem, and the principal is the only entity present in all four relations.

The same structural property has a direct privacy consequence. When the principal holds the reference frame, only the principal can correlate the identifiers it uses across its sovereign transactions. A custodial architecture, by holding state across multiple relations, makes such correlation available to the custodian by default; the relational architecture makes it structurally unavailable to any party other than the principal itself. This property is particularly enforceable in jurisdictions where data minimisation is embedded in data protection law: the architecture aligns with the regulatory requirement rather than depending on policy assertions to meet it. Profiling and correlation are not prevented by promise; they are prevented by design.

\paragraph{Capacity for choice qualifies the frame.} It remains to say why the autonomous principal, in the sense of Part~1 \cite{page2023distributed}, is the entity that can hold such a frame. The defining feature of the autonomous principal in Part~1 is \emph{capacity for choice}: the ability to decide whether to enter into relation, on what terms, and with what counterparty. This is precisely what qualifies an entity to hold a reference frame, because a reference frame is constituted by the choices made within it. A custodian holds records; a principal holds a frame. The distinction is structural rather than nominal: a custodian's records are intelligible to the extent that some external authority interprets them, while a principal's frame is intelligible to the extent that the principal can be recognised as the origin of choices that have causally ordered its history. The first depends on a hierarchy of interpretation; the second depends only on the principal's standing as a chooser.

\paragraph{Closing the structural argument.} Once these payoffs are taken together, transactional sovereignty ceases to be a normative preference for user empowerment and becomes a structural consequence. It is what survives the collapse of the custodial assumption (\S\ref{sec:newtonian}) once the regimes in which that assumption breaks (\S\ref{sec:fails}) become the regimes in which identity must routinely operate. The 2023 model of Part~1 was structurally required; what was missing in 2023 was the language, drawn from distributed systems and the causal structure of relativity, to say why. That language has become available through its first concrete applications, which we now present.
\begin{figure}[htbp]
    \centering
    \includegraphics[width=0.5\textwidth]{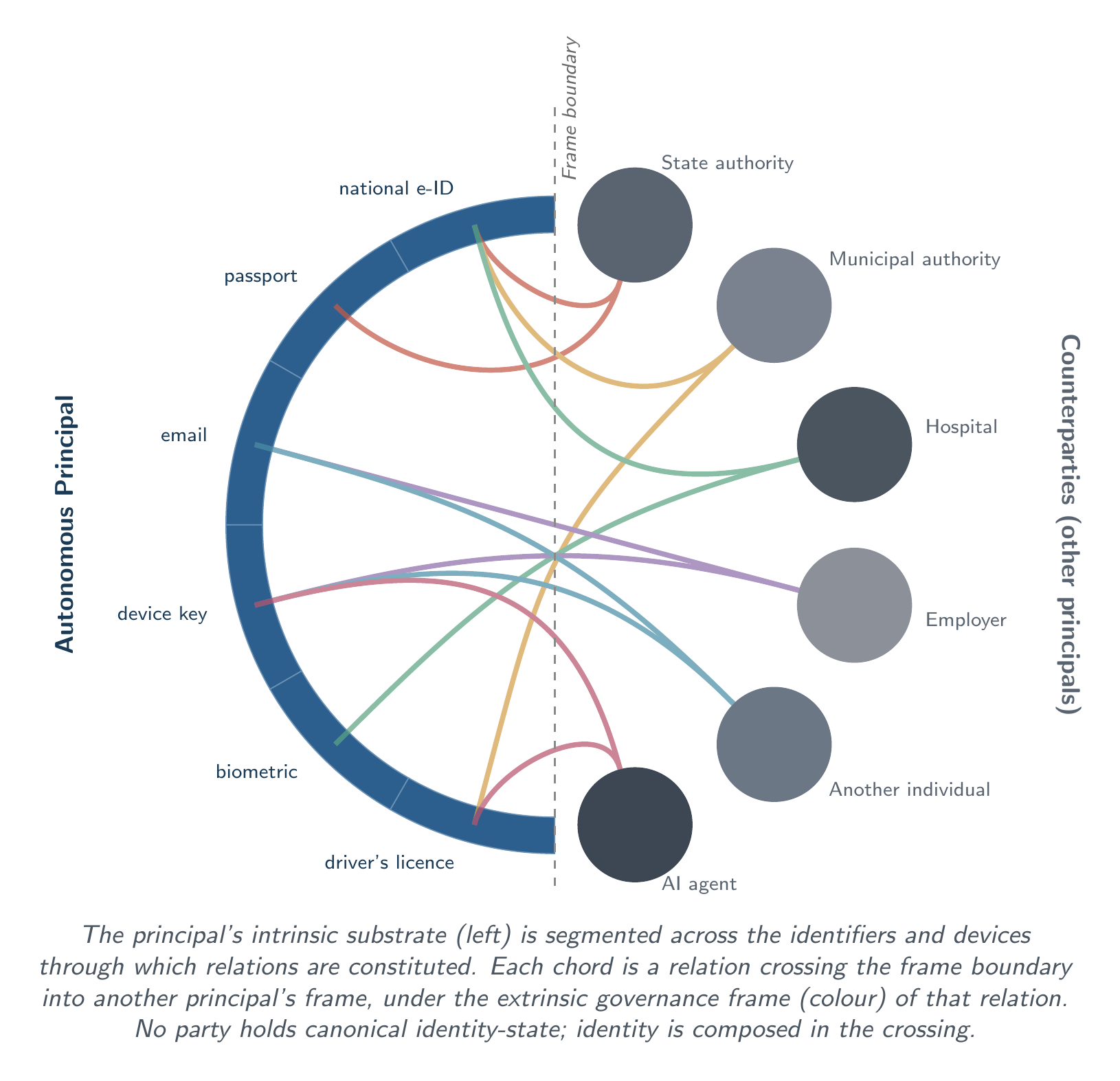}
    \caption{The relativistic picture, in compact form. The principal holds the reference frame from which its own digital identity is defined (left); other principals, each holding their own frame, appear opaque from outside (right). Identity is the relation that crosses the frame boundary, constituted under the extrinsic governance frame of that relation (colour).}
    \label{fig:relational}
\end{figure}
%
\section{What this is not}
\label{sec:not-this}

Section \ref{sec:newtonian} positioned this work as a continuation of the SSI programme rather than a critique of it. Two further distinctions are worth making explicit.

\textit{Relational identity is not blockchain-anchored identity.} A global ledger reintroduces absolute simultaneity by the back door, paying enormous coordination cost to manufacture a single global ordering of identity events. This is the consensus-protocol solution applied where it is not warranted: identity does not require a single global timeline, only causally ordered bilateral and multilateral exchanges \cite{lamport1978time,shapiro2011crdts}. Where global ordering is genuinely required --- registries of public authority, for instance --- consensus mechanisms remain appropriate. Where it is not, they impose the very Newtonian assumption we are trying to leave behind, now at the protocol layer rather than the institutional one.

\par
\textit{Relational identity is not user-controlled custody.} The shift we propose is not from \emph{the issuer holds your state} to \emph{you hold your state}. Holding remains a custodial relation; we are pointing past custody altogether. State, in the relational view, is not something held by any party; it is constituted between principals through causally ordered exchanges. The principal's role is not to be a better custodian of their own data but to be the reference frame from which a state of affairs is defined. This is a stronger claim than user-centricity, and a different one.

\par
\textit{Relational identity extends the SSI role triangle.} The issuer-holder-verifier model identifies the three structural roles of a credential exchange. What the model treats as properties of distinct parties --- one is the issuer, another the holder, a third the verifier --- the relational view treats as positions any autonomous principal can occupy, depending on the specifics of the transaction. A citizen is the holder of a passport issued by the state, the issuer of a tax declaration filed with the tax authority, and the verifier of a delivery driver's credential presented at the door. The same principal occupies all three roles across the relations in which it participates. The structural treatment of this generalisation is taken up in \S\ref{sec:composition}; we flag it here because it follows directly from the relocation of state, and is the move the SSI community is likely to recognise most readily as continuous with its own programme.

\par
\textit{Relational identity is not federated identity.} The most common engineering response to the multi-ecosystem problem is to introduce a single sign-on (SSO) provider, a passkey ecosystem, or a federation protocol that lets one custodian act as the identity authority for many --- the practitioner's reflex on first encountering the regimes of \S\ref{sec:fails}. These are real solutions to real engineering problems and we do not dispute their utility within the scope they address. We do dispute that they address the structural problem named in this paper. SSO and federation operate by introducing additional custodial infrastructure to compensate for the missing external frame; they preserve the Newtonian assumption rather than dissolving it, by elevating a single custodian to serve as the privileged vantage. The structural consequences follow directly: the custodian becomes a single point of failure for any principal who depends on it, and the architecture imposes a unifying frame on relations whose governance is plural by design --- citizen-to-state, patient-to-clinician, employee-to-employer, voter-to-municipality. The relational architecture takes the plurality as a structural feature and constitutes each relation under its own governance frame; federation, by contrast, manufactures a unifying frame that the underlying relations do not require. The deeper consequence is structural: a single external custodian, however well-engineered, is incompatible with digital sovereignty. A government cannot delegate to a foreign provider the frame from which its citizens' identity is defined; a corporation cannot delegate the frame from which its commercial relations are constituted; a natural person cannot delegate the frame from which its own life is read back. Each of these is a different face of the same structural fact: sovereignty over identity is not transferable to a custodian, because the frame is constitutive of the principal, not a service that can be procured from outside it.

\section{Operationalising relational identity}
\label{sec:technology}

\subsection{Primitives}
\label{sec:primitives}

\begin{figure}[htbp]
    \centering
    \includegraphics[width=0.95\textwidth]{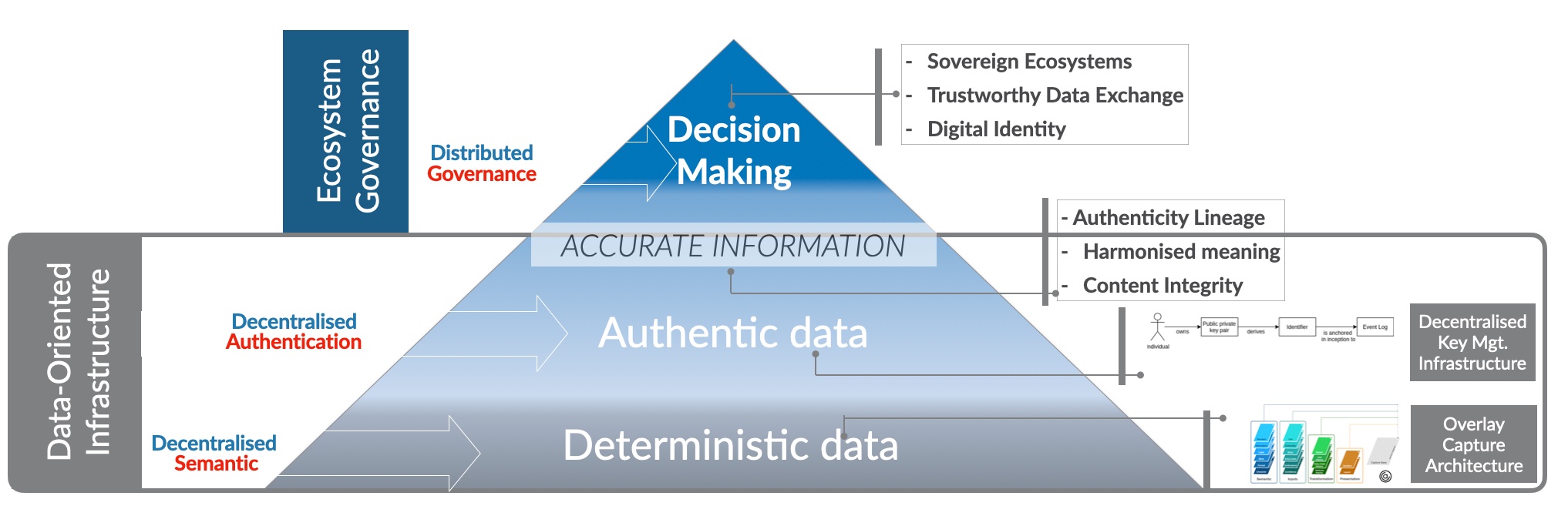}
    \caption{The relational accurate data stack. Decentralised Semantic and Decentralised Authentication form the data-oriented infrastructure, producing deterministic data and authentic data respectively. Their composition yields accurate information --- data whose authenticity lineage, harmonised meaning, and content integrity are verifiable end-to-end. Distributed Governance operates orthogonally across the stack, constituting the ecosystem in which decision-making (authorisation, consent, delegation) becomes meaningful. The three protocols are independent: each must be decentralised on its own terms, since a custodial dependency in any one reintroduces the Newtonian assumption (\S\ref{sec:newtonian}) through the back door.}
    \label{fig:stack}
\end{figure}

The architecture rests on three independent protocols, each addressing one dimension of the relational substrate identified in \S\ref{sec:argument}. \emph{Decentralised Authentication} secures the binding between a principal and the cryptographic material that lets that principal act under its own authority. \emph{Decentralised Semantic} secures the binding between data objects and the meaning under which they are exchanged. \emph{Distributed Governance}, as defined in Part~1 \cite{page2023distributed}, secures the binding between an ecosystem and the rules under which its principals interact. The three protocols are deliberately independent: each must be capable of decentralisation on its own terms, because dependency on a centralised provider in any one of them reintroduces custodial state through the back door, regardless of how relational the others are. The architecture is shown in Figure~\ref{fig:stack}.

\paragraph{Decentralised Authentication.} The authentication layer is provided by the Decentralised Key Management System (DKMS) developed by the Human Colossus Foundation \cite{dkms-hcf}, a system built on the KERI specification \cite{smith2019keri, toip-keri-spec}. The mechanism at the root of this layer is the self-certifying identifier --- an identifier derived directly from a public key, or from its content-addressed hash, such that the binding between the identifier and the controlling key material requires no external registry to attest. The concept is not new to KERI: it originates in Mazières's self-certifying pathnames, which moved key management outside the file system by letting the identifier carry its own public key \cite{mazieres2000sfs}. What KERI \cite{smith2019keri,toip-keri-spec} adds is the temporal dimension: a key-event log that records the identifier's control history, and a pre-rotation mechanism that cryptographically commits to the next key before it is needed, so that control provenance is end-verifiable across key rotations and not merely at a single moment. Control is established at inception, evolved through a pre-rotation mechanism that cryptographically commits to the next key before it is needed, and recorded in an append-only key event log that constitutes end-verifiable provenance of control. Delegation and multi-signature thresholds are first-class primitives. In the relational-identity stack, DKMS supplies the answer to \emph{which principal speaks here?} without reference to a privileged authority, and supplies it in a form that remains verifiable across jurisdictional and ecosystem boundaries.

\paragraph{Decentralised Semantic.} The semantic layer is provided by Overlays Capture Architecture (OCA) \cite{oca-hcf}, independently developed at the Foundation. OCA decomposes a data object into a stable \emph{capture base} --- the minimal structural definition of the object's attributes, bound to a self-addressing identifier derived from its content --- and a set of \emph{overlays} that carry contextual definitions (language, format, encoding, presentation, transformation, flagged attributes, validation). Because the capture base is content-addressed, any party holding a data object can verify that the schema under which it was issued has not changed; because overlays are linked rather than embedded, the same capture base can carry different contextual readings across jurisdictions, languages, or use cases without forking the underlying object. OCA supplies the answer to \emph{under what meaning is this data exchanged?} in a form that is composable, verifiable, and does not require a central registry to mediate semantic agreement.

\paragraph{Distributed Governance.} The governance layer is as defined in Part~1 \cite{page2023distributed}: an ecosystem is the unit of governance, constituted by autonomous principals operating under explicit and mutually verifiable rules of interaction. Where DKMS and OCA secure the data-oriented infrastructure, distributed governance secures the ecosystem-oriented one. It does not impose a global authority across ecosystems; it provides the primitives under which an ecosystem's rules are themselves the object of agreement among its principals, and under which inter-ecosystem composition is governed by the relations between principals rather than reconciliation between registries.

Taken together, the three protocols supply the substrate on which authorisation becomes meaningful. Authentic data (DKMS-secured) plus deterministic semantic (OCA-secured) yields what Figure~\ref{fig:stack} labels \emph{accurate information}: data whose authenticity lineage, harmonised meaning, and content integrity are verifiable by any party without recourse to a privileged custodian. Decision-making --- consent, delegation, authorisation --- can then take place on a substrate where the structural prerequisites for meaningful decisions are satisfied at the protocol layer rather than declared at the policy layer.

\subsection{Protocols}
\label{sec:protocols}

Having identified the three independent protocols and the substrate they jointly form, we describe the primitive operation through which a relational state is established between two principals, and the manner in which that state evolves causally over time.

\paragraph{The primitive relation: Out-of-Band Introduction.} A relational state is established when an autonomous principal performs an \emph{Out-Of-Band Introduction} (OOBI) with another autonomous principal \cite{smith2019keri}. The act consists in exchanging, through any channel the principals choose, the self-certifying identifiers and the locations from which their respective key event logs can be retrieved. From that exchange, each principal can independently verify the other's control provenance and establish a direct communication channel between them. No registry mediates the introduction; no authority registers the relation; the relation comes into existence as a fact between the two principals and is recorded in the causal log of each.

For the general reader it is worth dispelling any impression that the OOBI is an elaborate construction. In its simplest form a KERI OOBI is just a string --- a pairing of an identifier with a location at which to begin discovering information about it \cite{smith-oobi-spec}. It can be written as an ordinary URL:
\begin{quote}
\small\ttfamily
https://witness.example.org/oobi/\\
EHe5GYpz8Dyfb5Kk5KqJ\_iaThr1aRPjUz6r6\,...
\end{quote}
\noindent The trailing component is the principal's self-certifying identifier; the host is merely a hint as to where information about that identifier may be found. The string itself carries no authority and is not trusted: it only jump-starts discovery, and everything subsequently retrieved is verified in-band against the identifier's own key-event history. This is what makes the introducer disposable. The locator may be a witness, a watcher, an email, a QR code, or a slip of paper; once the identifier is known and its key-event history verified, the route by which it arrived no longer matters.

This act constitutes the primitive consensus on which all further relational state is built. In the terms of Part~1 \cite{page2023distributed}, it is the moment at which the confidentiality sphere of the relation is established: the principals agree, by the very act of OOBI, on whether and how they communicate --- privately, publicly, or under some defined intermediary regime --- and on the message definitions under which subsequent exchanges will be interpreted. The OOBI is therefore not merely an introduction in the cryptographic sense; it is the constitutive act of the relation, and the smallest unit of relational state from which everything else is built. 

This separation of introduction from authentication is not a novelty of digital systems. Humans have always distinguished the act of introduction --- the vouching that brings two parties into contact --- from the ongoing verification that subsequent messages come from the same party. A letter of introduction, a wax seal, a signet ring, the diplomatic presentation of credentials: each is an introduction mechanism whose verification depended on recognising a human-made mark. What KERI \cite{smith2019keri} contributes is not the separation itself but the nature of the verification layer: the identifier is cryptographically self-certifying, so that verification no longer depends on recognising a seal that a forger might reproduce, and the introducer becomes genuinely disposable --- once the OOBI has occurred, the relation runs on the cryptographic binding between the principals, not on the introducer's continued vouching. The architecture was, in a sense, always there in human practice; what was added was a verification layer that is non-repudiable and an introducer that the relation no longer depends on.

\paragraph{Fractal composition.} Upon this primitive relation more complex constructs emerge. Two principals in OOBI may invite a third under terms governed by their existing relation; an ecosystem is constituted when a set of principals enter into OOBI under a shared governance frame; cross-ecosystem composition occurs when principals from distinct ecosystems enter into bilateral or multilateral OOBI under terms negotiated between the ecosystems' governance layers. The structure is fractal in the technical sense: the same primitive (OOBI under explicit terms) operates at every scale, from the bilateral relation between two natural persons to the multilateral relation between institutions across jurisdictions. There is no privileged scale at which a custodian becomes necessary, because the primitive does not change as the scale increases.

\paragraph{Causal evolution.} Identity, in the relational view, is transactional: the state of any relation evolves through the causally ordered sequence of exchanges between the principals it constitutes. Two protocol-layer mechanisms enforce the causal path. First, DKMS guarantees that every exchange is signed under control material whose provenance is end-verifiable: the recipient of any message can establish, without reference to a third party, that the message was authored by the principal it claims and that the principal's control over its identifier has evolved through a verifiable sequence of key events. Second, OCA guarantees that every exchange refers to data objects whose schema and content integrity are independently verifiable: the recipient can establish, without reference to a third party, that the meaning under which the data is exchanged has not been altered between issuance and receipt.

The composition of these guarantees is what Figure~\ref{fig:stack} labels \emph{accurate information}. It is the precondition under which decision-making at the apex of the stack --- authorisation, consent, delegation, revocation --- can be meaningful. A consent granted under OCA-defined semantic with DKMS-authenticated provenance is a consent whose causal lineage can be reconstructed and verified at any later time by any party in the relation, without recourse to a custodian's record. A revocation propagates through the relation by the same mechanism that established it: as a causally ordered event signed under the revoking principal's evolved control, interpretable under the same semantic frame as the original grant. The propagation-lag regime of \S\ref{sec:fails} does not arise, because there is no registry to be lagged against; the truth of the relation is what the causal log between the principals records, and that log is by construction what each principal possesses.

\paragraph{The scope of consensus.} The architecture is deliberate in where it does and does not invoke consensus. Bilateral and multilateral relations between principals do not require global consensus: their truth is the truth of their own causal log, and global ordering is neither required nor meaningful. Where global ordering is genuinely needed --- for instance, in the publication of an ecosystem's governance frame, or in the public registry of an institution that exercises public authority --- consensus mechanisms (including, where appropriate, distributed ledgers) remain available, but they are confined to the scope at which they are warranted. This containment of consensus is a direct expression of the structural argument of \S\ref{sec:argument}: global ordering is the price paid to manufacture a classical reference frame, and it is paid only where such a frame is actually required, not as a default architectural assumption.
\par
 What determines the scope at which global ordering is warranted is, in practice, the ecosystem's governance framework: the normative structure that defines what the units of the ecosystem are and which acts within it require shared visibility across all participants. For a polity, this framework is its constitutional and legal order --- in the Swiss case, the federal constitution and the hierarchy of norms that descends from it, which sets the structural conditions under which an act of digital signature collection (\S\ref{sec:ecollecting}) requires recognition across distinct levels of authority. The architecture does not invent the scope of global ordering; it inherits it from the governance framework, and confines consensus to that scope rather than imposing it as a default.

\subsection{Composition across ecosystems}
\label{sec:composition}

The technical apparatus of \S\ref{sec:primitives} and \S\ref{sec:protocols} restates in the digital domain a structure that human societies have evolved over centuries of practice. An individual chooses whether to enter into relation with another; the other agrees or declines; out of such acts of mutual choice the institutions of civil and economic life are built. The principles under which a contract is formed, an association joined, a citizen recognised, or a purchase concluded are not novel inventions of the digital era; they are the long-standing protections that legitimate authority has accreted around the relational fact of human cooperation. The architecture we describe does not propose a new theory of association. It proposes that the digital re-instantiation of association should preserve the structural shape of its non-digital ancestor, in particular the placement of authority over the relation with the parties to it.

This is what \emph{transactional sovereignty}, as defined in Part~1 \cite{page2023distributed}, means in operation. A principal exercises sovereignty by deciding whether to enter into a relation, and on what terms; the counterparty exercises the same sovereignty by accepting, refusing, or counter-proposing. No prior register of permitted relations is required, because the relation comes into existence as a fact between the parties and is governed by the rules they agree to.

\paragraph{The fractal property in operation.} The structural claim of \S\ref{sec:protocols} --- that the same primitive operates at every scale --- is what makes cross-ecosystem composition possible. An ecosystem, in the sense of Part~1 \cite[Def.~2]{page2023distributed}, is a community of autonomous principals bound by a legitimate authority and constituted by the rules of interaction its principals have adopted. Because the ecosystem is itself a unit of governance, sovereign within its own perimeter, it can stand on either side of an OOBI: a principal can enter into relation with an ecosystem under the same protocol by which two principals enter into relation with one another. Two concrete cases illustrate the symmetry.

Consider first passport issuance in a federal state. Three legitimate authorities are involved in the same act, each at a distinct structural level. The federal state, as the highest-order ecosystem, holds the constitutional mandate to recognise its citizens internationally. A canton (or comparable sub-national authority), as a constituent ecosystem within the federation, exercises the operational authority to issue passports to its residents under the federal mandate. The citizen, as an autonomous principal, is the subject whose identity is attested by the resulting credential and who then exercises that credential in further relations --- crossing borders, opening bank accounts, registering with foreign authorities --- in which none of the three issuing layers is the custodian of the relation. The same protocol operates at every level: the federation enters into relation with the canton under the terms its constitution defines; the canton enters into relation with the citizen under the terms federal and cantonal law jointly specify; the citizen enters into relation with the foreign authority under the terms the receiving ecosystem accepts. No layer holds the canonical state of any other; each layer holds the relations to which it is a party. A concrete deployment of this pattern in Swiss federalism is developed in \S\ref{sec:ecollecting}.

Consider second a customer purchasing goods from a shop. The shop, as an ecosystem, has its own internal governance --- its staff, its policies, its terms of trade. The customer is not a member of that ecosystem; the customer is an external autonomous principal entering into a transactional relation with it. Yet the protocol that governs the transaction is the same one that governs the citizen-state relation: a mutually agreed entry into relation, under terms the parties have established, recorded as a causally ordered exchange between them.

\paragraph{Ecosystem-to-ecosystem composition.} The same protocol extends a level higher. When company A purchases goods from company B, two ecosystems --- each with its own internal governance --- enter into a peer-to-peer relation. The relation may be bilateral, governed by the contract between them; or it may be cast into a higher-order ecosystem, such as a trade association, whose governance frame is the object of agreement among its member ecosystems. In each case the structural pattern is identical: a relation is constituted by the parties to it, under terms they have adopted, with no requirement that a higher authority above them mediate the canonical state of either.

\paragraph{Variety of principals and ecosystems.} The architecture admits the variety of principals defined in Part~1 \cite[Def.~1.1--1.3]{page2023distributed} --- natural persons, legal persons, and governmental entities. Moreover, the variety of ecosystems --- which may include non-autonomous agents as constituents --- spans the complexity levels Part~1 characterises explicitly. A given principal may participate in many ecosystems concurrently, each governing a different aspect of its activity, and may enter into transactional relations with ecosystems to which it does not belong. The complex web of overlapping ecosystems that constitutes a society translates directly into the digital domain: not as a hierarchy whose apex is a single custodian, but as a graph of relations whose vertices are principals and ecosystems and whose edges are governed by the protocols of \S\ref{sec:protocols}.

\paragraph{Why custody cannot do this.} A custodial architecture applied to this regime need not maintain a single central registry of every principal --- public-key infrastructure shows that custodial trust can be distributed operationally, through hierarchical delegation from a root of trust down through intermediate authorities to end entities, with no global list of all principals anywhere in the system. What custody cannot escape is the dependency that this delegation creates: every principal's standing ultimately descends from, and depends upon, a shared root frame. The operations are distributed; the trust is not. This is the structural vulnerability. Compromise of the root --- or of any sufficiently high node in the delegation hierarchy --- cascades to everything that depends on it. The 2011 compromise of the Dutch certificate authority DigiNotar is the canonical demonstration: an intruder who gained control of the CA infrastructure issued more than 500 fraudulent certificates for major domains, one of which was used to intercept the communications of an estimated 300,000 users \cite{foxit2012blacktulip}; because DigiNotar was also the certificate authority for the Dutch government, and because browsers could only respond by removing the DigiNotar root from their trust stores entirely, the compromise collapsed every certificate the authority had issued --- legitimate and fraudulent alike --- and the company entered bankruptcy within weeks \cite{cfr-diginotar}. The Comodo compromise earlier the same year showed the same structural pattern \cite{comodo2011}; these are not isolated failures of particular authorities but recurrent expressions of a single architectural fact. The shared root is precisely the privileged frame that \S\ref{sec:frame} identified as the hidden assumption of custodial identity: a single vantage on which the legitimacy of every principal's standing ultimately depends. The relational architecture has no such root. Each principal's identifier is self-certifying, rooted in the principal's own key material rather than in a shared external authority (\S\ref{sec:primitives}); trust is established relationally, between principals, under governance that is itself part of the relation. There is no node whose compromise cascades across the system, because there is no node on which the system's trust globally depends.
\par
The mechanism is visible in current production systems. Single sign-on and passkey ecosystems both work by binding a principal's identity to a credential held by an identity provider, with that provider attesting on the principal's behalf to every relying party. This is engineering that succeeds at what it was designed to do --- it reduces credential proliferation, raises the security floor against many common attacks, and is widely deployed for good reason. It fails, however, at the structural property at issue here: the principal's continued participation in their relations is contingent on the identity provider's continued willingness and ability to attest, an outage at the provider becomes an outage across every relying party simultaneously, a compromise at the provider becomes a compromise across every relying party simultaneously, and a principal who leaves the provider's ecosystem cannot carry the relations the provider brokered. The provider has captured the principal's relations in the precise sense that no part of the architecture allows the principal to depart with them intact. Relational identity, by contrast, locates the binding in the principal's own substrate: relations remain valid as long as the principal's keys remain in control, regardless of which provider is online or which ecosystem the principal continues to participate in.

\paragraph{Relational identity extends the SSI roles.} A note worth emphasising for readers from the SSI community. The issuer-holder-verifier model identifies the three structural roles of a credential exchange. What the model treats as properties of distinct parties --- one is the issuer, another the holder, a third the verifier --- the relational view treats as positions any autonomous principal can occupy depending on the context and transaction. A patient is the holder of a credential from their hospital, the issuer of a credential to a research consortium that joins their cohort, and the verifier of a credential presented by a clinician requesting access. The same principal occupies all three roles across the relations in which it participates, and the architecture must support this without architectural discontinuity. The relational-identity model does so natively, because the roles are properties of the relation, not of the principal.

\subsection*{Transition to applications}
The three subsections above describe the architecture in full: the primitives on which it rests (\S\ref{sec:primitives}), the protocol mechanics by which a relational state is constituted and evolved (\S\ref{sec:protocols}), and the manner in which the same primitives compose across principals and ecosystems of arbitrary complexity (\S\ref{sec:composition}). What remains is to demonstrate that the architecture meets the structural requirements of \S\ref{sec:argument} in concrete deployment. The five applications of \S\ref{sec:applications} are each presented to expose a different facet of the structural argument: guardianship, where the holder relation itself must be constituted under governance; personalised medicine research, where principals interact across conflicting jurisdictional frames; the Swiss federal e-collecting case, where a single sovereign act is constituted across multiple distinct constitutional authorities none of which can act as custodian for the others; patient-centric data exchange, where the requirements of the principal community have been articulated independently and the architecture is the response;  and the Polish \emph{Truth on the Web} initiative, where the portability of authenticated public information beyond the custodian's domain is the structural property at stake, with an institutional path toward production deployment within an existing national digital infrastructure.

\section{Applications}
\label{sec:applications}

We present five applications, chosen because each instantiates a structurally distinct regime in which the Newtonian assumption fails. Table~\ref{tab:regime-application} (page~\pageref{tab:regime-application}) summarises the mapping; the prose that follows unpacks each case in turn. Guardianship (\S\ref{sec:guardianship}) is the constituting regime: the holder relation itself must be governed. Personalised medicine research (\S\ref{sec:nextgen}) is one sub-case of the cross-jurisdiction regime, in which multiple distinct legitimate authorities, each sovereign within its own jurisdiction, must converge on a shared governance frame for a specific purpose. Federated electronic signature collection (\S\ref{sec:ecollecting}) is the other sub-case of the cross-jurisdiction regime, in which a single undisputed legitimate authority is administered by a plurality of autonomous administrative entities each exercising distinct constitutional mandates over the same act. Patient-centric data exchange (\S\ref{sec:mpne}) is one sub-case of the propagation-lag regime in which institutional data diverges from the operative truth of clinical and consent state; the community of principals has explicitly articulated its structural requirements. The Polish \emph{Truth on the Web} project (\S\ref{sec:truthontheweb}) is the other sub-case of propagation-lag, where authenticity must survive beyond the custodian's domain as content propagates through the digital public sphere. Together the five cover the failure modes of \S\ref{sec:fails} and demonstrate that one architecture resolves all of them, across a range of deployment maturity from articulated requirement through prototype to operational integration with existing projects.
\begin{table}[ht]
\centering
\small\sffamily
\caption{The structural argument of \S\ref{sec:argument} mapped to the applications of \S\ref{sec:applications}. Each application instantiates a regime in which the Newtonian assumption fails. Where a regime admits multiple structurally distinct sub-cases, both are shown; together the applications cover the failure modes of \S\ref{sec:fails}.}
\label{tab:regime-application}
\begin{tabular}{p{3.4cm}p{6cm}p{4.4cm}}
\toprule
\textbf{Regime (\S\ref{sec:fails})} & \textbf{Structural property at stake} & \textbf{Application} \\
\midrule
\textbf{Constituting regime} &
The holder relation itself must be constituted under governance; the holder cannot be presupposed. &
Guardianship (\S\ref{sec:guardianship}) \\
\addlinespace
\multicolumn{2}{l}{\textbf{Cross-jurisdiction regime:}}& \\
\textit{-- plural legitimate authorities constituting a common frame} &
Multiple distinct legitimate authorities, each sovereign within its own jurisdiction, must converge on a shared governance frame for a specific purpose --- here the cross-border exchange of sensitive data. &
Personalised medicine research (\S\ref{sec:nextgen}) \\
\textit{-- multiple administrative entities under one constitution} &
A single undisputed legitimate authority (the federal constitution) is administered by a plurality of autonomous administrative entities (Communes, Cantons, Federal Chancellery), each exercising distinct constitutional mandates over the same act. &
Federated electronic signature collection (\S\ref{sec:ecollecting}) \\
\addlinespace
\multicolumn{2}{l}{\textbf{Propagation-lag regime:}} & \\
\textit{-- institutional data} &
The custodian's record diverges from the operative truth of clinical and consent state as it propagates across institutional boundaries within an ecosystem. &
Patient-centric data exchange (\S\ref{sec:mpne}) \\
\textit{-- public information} &
Authenticity must survive beyond the custodian's domain as content propagates through the digital public sphere. &
\emph{Truth on the Web} (\S\ref{sec:truthontheweb}) \\
\addlinespace
\textbf{Agentic regime} &
The principal acting in a relation is an agent whose state is constituted by the delegation chain that traces back to an autonomous principal. &
Forthcoming work, outlined in \S\ref{sec:conclusion} \\
\bottomrule
\end{tabular}
\end{table}

\subsection{Guardianship and the limits of the holder model}
\label{sec:guardianship}

In December 2019, the Sovrin Foundation Guardianship Task Force, mandated by the Sovrin Governance Working Group, published \emph{On Guardianship in Self-Sovereign Identity} \cite{sovrin2019guardianship}, the first systematic treatment of guardianship in the SSI context. The paper opens with a structural observation: as powerful as SSI is, the issuer-holder-verifier model is directly usable only by subjects who have both digital access and the legal and mental capacity to act for themselves. For everyone else --- the refugee without a device, the child, the elderly person living with dementia, the patient under anaesthesia, the deceased estate --- another party must act as digital guardian. The paper estimated that the subjects requiring guardianship constitute a substantial fraction of the global population. Subsequent work by the Sovrin Guardianship Working Group developed implementation guidelines and technical requirements \cite{sovrin2021guardianship-impl,sovrin2023guardianship-v2}, and the topic has continued to develop across the broader digital identity community, including the Decentralized Identity Foundation, the W3C Verifiable Credentials and DID working groups, and the Trust over IP Foundation \cite{dif2024guardianship,w3c-did,w3c-vc}\footnote{the Sovrin Foundation was dissolved in 2025 \cite{sovrin2024dissolution}}.

What the working group correctly identified, and what subsequent work has not closed, is that guardianship is not a credential type but a relation. The guardian acts for the dependent under conditions that themselves require governance: the scope of authority granted, the conditions under which it terminates, the mechanisms by which the relation is established, observed, and dissolved as the dependent acquires or loses capacity. The issuer-holder-verifier model can express guardian credentials, but it cannot natively express the guardianship relation itself, because the model presupposes a holder who is the subject's own locus of control. When the holder must be constituted by an external relation under governance, the model has no place to put that relation: the relation precedes the holder, and the architecture begins downstream of where the problem is.

The structural argument of \S\ref{sec:argument} gives this problem a natural home. If state is relational rather than held, then guardianship is not an exceptional case requiring special architectural treatment. It is what relational identity always already is: a state of affairs constituted between principals under explicit governance, with causally ordered exchanges defining the relation's establishment, evolution, and termination. The dependent's reference frame is partially or fully expressed through the guardian's, under terms that are themselves part of the relation and themselves subject to governance. The transition into and out of self-custody as the dependent acquires or loses capacity is not a discontinuous architectural event; it is a continuous modulation of the relation's parameters.

The technology described in \S\ref{sec:technology} treats this case as a primitive rather than an extension. The point is best made by walking through one of the use cases set out in \cite{sovrin2019guardianship}. Take the refugee case, Mya. Under the issuer-holder-verifier model her trajectory from no capacity toward self-determination is punctuated by architectural events --- her registration in a refugee camp, the assignment of guardianship to a humanitarian agency, family reunification, the eventual acquisition of a device --- each of which the custodial model must absorb as a discontinuity. The Newtonian assumption compounds the problem: multiple organisations concurrently hold custodial state for the same refugee, often inconsistent, and the guardianship state itself becomes an additional layer of complexity on top.

Consider Mya as a five or six-year-old in poor health, arriving alone in a refugee camp. The relation begins with the establishment of a guardianship bond between Mya (the dependent) and the humanitarian agency (the guardian). This first OOBI is made under the governance of the agency, designed to extend full protection to the refugee against the threats of the surrounding environment. OOBI as an introduction mechanism in the sense of \S\ref{sec:protocols} will most probably not be digital. Mya has no capacity in any operative sense; her identity, as a relationship between Mya as a person and her context, must be reconstructed. These relations are an OOBIs within a governance frame --- a registration process under explicit institutional rules --- not a credential.

Mya's journey to self-determined digital status will involve multiple agencies providing humanitarian services: food, medical care, transfer to safer locations, family reunification. Her parents are lost or dead, and an unofficial foster family becomes a temporary guardian simply because they recognised her from their home village --- not because a formal identity relation existed. The constitution of a custodial identity state under the Newtonian assumption is both operationally complex and dangerous in this regime. It requires a high level of interoperability between heterogeneous organisations operating in unstable real-world conditions \cite[\S1.1]{page2023distributed}. And even where a custodial state of identity is achieved, it becomes the attractor of the very threats from which the refugee is attempting to escape. The January 2022 attack on the ICRC's Restoring Family Links data, in which the personal data of more than 515,000 vulnerable people held by an external contractor in Switzerland was compromised \cite{icrc2022rfl}, is the case study of what happens when a custodial reduction of humanitarian identity meets a determined adversary.

In the relational approach, with identity expressed as a causal log of events, the requirement to centralise identity information is reduced. It is replaced by a structure in which the refugee is the principal whose digital capacity evolves in parallel to the re-identification process. The first electronic registration in an agency system may use Mya's biometric as the only binding to her physical person. Records may be accessible to the temporary foster family. From the outset, each of the relations --- Mya--agency, Mya--guardian, guardian--agency --- is established through the same OOBI-plus-governance protocol; the first OOBIs may be verbal or paper-based, with no electronic record at all. OOBI's are From no device and no capacity toward self-determination and technological independence, many years may pass; the only thing required throughout is the causal lineage of identity state within Mya's own frame. Capacity is recovered through a continuous governance under that frame --- the frame of Mya as autonomous principal --- despite the discontinuity of the systems architectures she encounters along the way.

The same structural treatment generalises to AI agents in a direction the guardianship literature has not yet fully addressed. Every autonomous agent is, in a structural sense, a guardianship case in reverse: a principal acting through a non-principal that cannot hold its own reference frame. The agent operates on behalf of a principal under terms that must be governed --- scope of authority, conditions of action, mechanisms of observation and revocation. This is the same relational structure as classical guardianship, with the roles of capacity-holder and dependent re-allocated. An architecture that handles human guardianship and agent delegation under the same primitives is, we claim, the natural one for the regime that the agent-governance literature is currently grappling with from the outside \cite{chaffer2024decentralized,ruan2026agentcity}.

\medskip
\noindent\textit{Author's note.} The first author was a contributing member of the \emph{Sovrin Guardianship Task Force} \cite{sovrin2019guardianship}. The problem was clear at the time of writing, June to November 2019, and the community's analysis of it remains a reference. What was not yet available was a structural account of why the issuer-holder-verifier model could not close the problem within its own terms, nor an architecture in which the closure was natural rather than bolted on. The argument and technology presented above are offered in that spirit.

\subsection{Personalised medicine: the researcher as principal}
\label{sec:nextgen}

The 4-years EU Horizon Europe project NextGen \cite{nextgen-cordis} develops, for personalised cardiovascular medicine, tools and infrastructure for multi-modal data integration in an ecosystem of clinical research centres and universities located in the European Union, the United Kingdom, and the United States, with the support of SMEs and civil-society organisations. The technical challenge is widely understood: integrate genomic, imaging, clinical, and other required modalities across research workflows that span multiple regulatory and governance regimes. As the recent state-of-the-art review of cardiovascular genomic and precision medicine emphasises \cite{chahal2025interop}, the integration of phenotypic data with multi-omics layers (genomics, transcriptomics, proteomics, metabolomics) presents a two-fold structural challenge: harmonising data across different technologies, and consolidating data for the same technology obtained from different centres using different methods --- problems that compound when the integration crosses jurisdictional boundaries. The governance challenge is less widely understood, particularly from the researchers' standpoint: despite legitimate research questions, they face a heavy regulatory framework for the secondary use of health data, of which the European Health Data Space \cite{ehds-regulation} is the most visible current expression. This is where current solutions based on custodial identity fall short, and where the relational-identity argument has direct purchase: the governance burden is not an artefact of regulation but a structural consequence of architectures that require each cross-jurisdictional act to be mediated through a custodial frame, where the relational architecture would constitute the act under its own native governance.

In the conventional framing, the patient is the subject whose data is governed and the researcher is the agent who accesses it. The custodial-IAM treatment of this scenario places the canonical state of the patient with an institutional custodian (the hospital, the national registry, the federated query infrastructure) and the canonical state of the researcher's authorisation with another (the institution, the ethics board, the data access committee). Cross-jurisdictional research must then reconcile these custodians against one another --- the cross-jurisdictional regime our \S\ref{sec:fails} diagnosed as structurally inapplicable to custodial state.

Privacy, or the confidentiality of professional interactions, is both personal and contextual. The regulations around privacy reflect the protection that a community expects when it comes to information exchange, and the GDPR reflects the EU's articulation of this expectation for personal data. The conceptual seed of the GDPR can be traced to the right of informational self-determination established by the German Federal Constitutional Court in its 1983 Census Judgement \cite{schlink1986volkszaehlungsurteil}, itself grounded in a more basic right to the free development of the person \cite{rouvroy2009right}. The link to the autonomous principal as defined in Part~1 is direct: informational self-determination is the legal articulation of the structural property the architecture treats as primitive. Under the Newtonian assumption, privacy-by-design is applied to the systems that act as custodians of identity. NextGen researches a different architecture. The project's tooling can be classified in three families, each contributing to privacy-by-design under this architecture: federated machine-learning pipelines that let local centres compute on data without centralising it; open-source, vendor-independent computation-intensive pipelines that protect institutions from the lock-in that proprietary platforms impose; and --- the line most relevant to the present paper --- a data-management architecture implementing the protocols of \S\ref{sec:technology} on a data-oriented substrate, detailed in the next paragraph.

The NextGen data architecture adopts a local-first, researcher-centric, governance-embedded design that moves toward the relational view: data remains under local control, governance is constituted within the architecture rather than overlaid as paper-based data agreements, and the researcher is equipped to expose and discover datasets while remaining within the rules of the local research organisation. The researcher is not merely an agent operating under institutional permission; the researcher is a principal whose research relation with each patient, or with each cohort under each consent regime, is itself the object of governance. The architecture is exercised through two operational primitives whose composition follows the relational pattern. \emph{Onboarding}: an out-of-band invocation (OOBI) connects a NextGen participant to the project's resources; DKMS supplies authentication; the governance defines what can be done with the identifiers and uses the OCA to describe the meaning of roles, credentials and any other data structure. The ecosystem states what it needs; local participants supply what they can; control remains distributed at both ends. \emph{Discovery}: queries across the federated catalogue are distributed to local nodes operating under potentially distinct jurisdictions. OCA carries the schema and semantics --- the verifiable definitional layer, together with simple transformations such as unit conversion that the overlay structure supports directly --- so that a node can reason over the meaning of what it holds without exposing the underlying records. Where a query requires data spanning several modalities, NextGen composes the relevant types into a Multimodal Integration Object (MMIO) \cite{nextgen-mmio}: a self-contained data object that binds the different modalities OCA schemas content to their policies, agreements, and carries its own integrity and authenticity through append-only logs and cryptographic identifiers, and can be shared across systems without moving the underlying data or shedding its governance rules. The local-first control is exercised through governance, under which each node defines its own semantics, states what it means by them, and decides what may be revealed in which context. DKMS supplies the resolution of the public key bound to each identifier, establishing that the querying principal and the queried node are genuinely the controllers of the identifiers they present; this cryptographic proof of control is necessary but not sufficient for authentication, which is completed only when governance establishes who the identifier is admitted to be in this ecosystem and what it is authorised to do. The MMIO is, in this sense, a concrete instantiation of the composition principle of \S\ref{sec:composition}: a single object in which OCA's verifiable meaning, the key-event integrity of the authentication layer, and the binding governance rules are composed into a portable, auditable whole that returns lawful responses to cross-jurisdictional queries --- with the FAIR principles \cite{wilkinson2016fair} emerging as a structural property of the architecture rather than as a separate compliance layer.

NextGen also researches how an ecosystem built on this architecture can interoperate with ecosystems still operating under custodial assumptions --- the blueprint for how researcher authorisation propagates across jurisdictions without custodial reconciliation, how consent revocation propagates causally rather than through registry synchronisation, and how alignment between the architecture and instruments such as the 1+ Million Genomes Initiative and the EHDS looks in practice.

The NextGen case demonstrates that the relational architecture handles the cross-jurisdiction regime as it appears in international research collaboration: multiple legitimate authorities, each sovereign within its own jurisdiction, converging on a shared governance frame for a specific purpose. The same regime presents a structurally distinct face when the multiplicity lies in the administration of a single legitimate authority rather than across legitimate authorities themselves. The next subsection considers that case in the form of federated electronic signature collection within a single constitutional order.

\subsection{Federated electronic signature collection: the Swiss E-Collecting case}
\label{sec:ecollecting}

In 2025, the Swiss Federal Chancellery launched a participative process to examine \emph{e-collecting} --- the collection of citizen signatures supporting popular initiatives and federal referenda, a core political right anchored in chapter 2 of the Federal Constitution \cite[art.138-142]{swiss-constitution-139}. The impetus was a series of frauds discovered in 2024, in which commercial signature-collection firms were found to have falsified signatures, prompting criminal complaints by the Federal Chancellery and a measurable erosion of public trust in one of the central instruments of Swiss direct democracy \cite{rts2024fraud}. In a November 2024 report \cite{cf2024ecollecting}, the Federal Council made a structural observation directly relevant to the architecture proposed here: that the participation of several mutually independent actors in the process fosters reciprocal control, that the decision on whether a popular request succeeds rests on the work of multiple distinct responsible parties --- the voter-eligibility attestation services, the Federal Chancellery, and the committees --- and that the decentralisation of these operations prevents localised malfunctions from propagating across the whole process. In autumn 2025, a hackathon was organised to frame technological solutions that could bring the process into the digital domain without forfeiting the structural protections that Swiss direct democracy depends on \cite{swiss-ecollecting-2025}. The authors participated with a proposal; during the hackathon, other participants joined the effort to form Team~8, which proposed an architecture built on the protocols of \S\ref{sec:protocols}. The broader programme is described in \cite{page2025evoting}.

The cross-jurisdiction regime is structurally distinct from the three preceding applications. It is not cross-jurisdictional in the inter-state sense: the entire process takes place within Swiss federal law. It is not propagation-lag-dominated: the act of signing is discrete and bounded in time; the change in citizen's commune during a signature collection period is a marginal factor. Yet it cannot be solved by a custodial architecture, because Swiss direct democracy distributes legitimate authority across four distinct layers \cite{cf2024ecollecting}. The \emph{Comité} initiates and authenticates the initiative and mandates collectors. The \emph{Chancellerie f\'ed\'erale} verifies the initiative's admissibility, counts the certified signatures, and issues the confirmation of success. The \emph{Cantons} and \emph{Communes} verify the eligibility of each signatory and certify that the signatures collected in their jurisdiction satisfy the constitutional requirements. As pointed out explicitely in  \cite[section 1.4.6]{cf2024ecollecting}, each of these authorities exercises a distinct constitutional mandate; none has standing to act as custodian for the others; the Swiss federal order is structurally built to refuse the centralisation that would make custodial reconciliation possible.

A simple custodial e-collecting platform would operate in tension with both of these reference points. A platform architecture aggregates in a single technical locus the record of which initiatives each citizen has chosen to support --- a record that the paper process is designed to be spread across actors legally bound to their separate duties.  The commune sees only the signatures it has verified and the Chancellerie sees only the verified count, and an ephemeral access to the signatories with the legal obligation to destroy them upon their confirmation of approval/rejection of the initiative. The Federal Council's own observation is that the decentralisation of operations across mutually independent actors is what prevents localised malfunctions from propagating and what sustains reciprocal control; a custodial architecture, by centralising the canonical state, removes precisely the decentralisation the report identifies as protective. The same centralisation stands in structural tension with the constitutional distribution of authority across the Comité, the Chancellerie, the Cantons, and the Communes, none of which the paper process places in possession of a federation-wide record. A custodial platform that honoured these constraints would therefore have to re-introduce, in software, the decentralisation its own architecture removes --- reconstructing the separation of authorities and the absence of a central record as features bolted onto a system whose default is their opposite. This is the source of the complexity: the process operates in a regime outside the Newtonian assumption, and a custodial architecture must work against its own grain to honour that regime. The structural alternative is to adopt an architecture whose default already matches the constitutional structure, so that no participant ever holds canonical state across the federation.

The architecture proposed by Team~8 addresses this not by replacing the central platform with a better one, but by preserving the constitutional split of responsibility across the distinct governances and using open protocols only to mediate the crossings between them, according to the rules each governance defines. The open protocols are the means, not the solution: what makes it possible to surpass a central platform is the clean separation of the Comité, the Chancellerie, the Cantons, and the Communes into distinct governance domains, each retaining authority over its own part of the process, with the protocols carrying the verifiable state across the boundaries between them under the constraints each domain imposes. Within this structure, the three primitives play defined roles. DKMS supplies the resolution of the citizen's self-certifying identifier, establishing control provenance that is end-verifiable without reference to a central registry --- the cryptographic proof of control that, as in \S\ref{sec:nextgen}, governance must then complete by establishing eligibility and admission. The Commune, after verifying eligibility through one of several identity providers (paper-based signature, SwissPass, Swiss e-ID, passport, SwissSign, or others, as accessibility and cantonal practice dictate), issues a one-time certificate that authorises a single signature on a specified initiative. Anonymity holds between the citizen and every party other than the verifying Commune; linkability is confined to the minimum required for eligibility verification and is dissolved thereafter. OCA secures the integrity and harmonisation of the initiative text and supporting materials across the four official languages and across accessibility-enhanced presentations (translations for inclusion, alternative formats for the visually impaired, simplified language for cognitive accessibility). Distributed governance secures the rules under which the Comité, the Chancellerie, the Cantons, and the Communes interact, without imposing a central platform that would stand above them. The architecture diagram and sequence diagrams documenting the relevant flows are presented in \cite{swiss-ecollecting-2025}.

\paragraph{The architecture is more complex than a platform, and necessarily so.} \textit{The reflections that follow are the authors' alone and do not represent the positions of Team~8 collaborators or their institutions.} A reception worth addressing directly: at the conclusion of the hackathon, the proposed approach was neither accepted nor rejected. Among the responses recorded was the observation that the architecture appeared \emph{complex} relative to centralised alternatives. The observation is correct on its face and worth examining structurally. A centralised platform is simpler because it ignores the problem that Swiss federalism is built to solve. The complexity of the relational-identity architecture is not gratuitous; it is the operational expression of the constitutional structure the architecture is designed to preserve. What "removing the complexity" would actually mean in this case is therefore worth stating plainly: removing the complexity is equivalent to removing the federalism, the separation of authorities, the anonymity-by-design that Swiss direct democracy has historically maintained. The relevant comparison is therefore not between a simple platform and a complex protocol stack; it is between two distinct architectures that produce different constitutional artifacts, and the choice between them is a choice about what is digitised --- the process or its constitutional shape. This paper exists in part to make that distinction visible, so that future deliberations on civic digital infrastructure can be conducted against the right comparison.

\paragraph{What the case shows.} The structural payoff of \S\ref{sec:argument} is most clearly visible here. The fractal claim of \S\ref{sec:protocols} --- that the same protocol operates from bilateral relations up through multi-layer ecosystems --- is instantiated concretely: the citizen-Commune relation, the Commune-Canton relation, the Canton-Chancellerie relation, and the Chancellerie-Comité relation are each constituted under the same primitives. No layer holds canonical state for any other; each holds the relations to which it is a party; the constitutional shape of Swiss direct democracy is preserved in its digital re-instantiation. That this preservation is more complex than a central platform reflects the irreducible complexity of the constitutional arrangement, not a failing of the architecture. The architecture was presented at the Federal Chancellery's E-Collecting Hackathon~2025; the deliberation on its further development belongs to the relevant Swiss institutions.

\subsection{Patient-centric data exchange: the MPNE consensus in practice}
\label{sec:mpne}
In early 2024, the Melanoma Patient Network Europe (MPNE) convened \emph{MPNEconsensus 2024} at the Fraunhofer Institute in Berlin, supported by the EU funded iToBoS project \cite{itobos-cordis}, and produced ten consensus statements on data, AI, and data-dependent business models in healthcare \cite{mpne2024consensus}. The statements are the result of a structured deliberation among melanoma patient advocates, reached by the patient community itself; project partners, including the Human Colossus Foundation, of which the authors are members, were present to provide technical expertise on request, but the statements are the community's own.

Several of the statements bear directly on the structural argument of this paper. The community holds that health data is a common good, and that it is not acceptable to extract value for a few while leaving patients and society to carry the risks (statement~4). It calls for \emph{zero-trust environments beyond the reach of single parties, institutions, or governments}, particularly for genomic data (statement~6). And it is explicit that technology cannot self-regulate, so that control must also be exercised through law --- hard guardrails, real-time monitoring, and enforcement with genuine consequences (statement~8). Read together, these statements describe a setting in which no single custodian --- whether a company, an institution, or a government --- can be the privileged locus from which the legitimacy of a data exchange is determined, and in which the architecture must be paired with external legal enforcement rather than substituting for it.

Read against the argument of this paper, the MPNE consensus is an articulation, by a patient community deliberating independently, of requirements that the custodial assumption cannot structurally satisfy. A custodial architecture places the canonical state of patient data with an institutional custodian; statement~6's demand for an environment beyond the reach of any single party is precisely the demand that no such custodian exist. The relational architecture meets this architectural requirement directly: the patient is the reference frame from which the legitimacy of any exchange is defined, and the truth of an exchange is determined by what the relation between principals authorises, expressed in causally ordered and verifiable form. The community's parallel insistence on legal enforcement (statement~8) is not in tension with this; it reflects the same recognition that runs through \S\ref{sec:technology}, that governance is constitutive of relations and is not supplied by the technical substrate alone. The architecture provides the structure within which enforceable governance operates; it does not replace the law the community rightly demands.

The consensus did not begin from a technical proposal. It began in 2022 as a Design Thinking process applying Service Design concepts to capture patient need and interest in an imaginary  application \emph{"Gillyweed"}, sketched by an MPNE advocate to dramatise what the European Health Data Space ought to deliver for patients if one started from patient need rather than from implementation constraints \cite{spurrier2022sync}. Worked through in a series of community workshops \cite{mpne-gillyweed}, Gillyweed became the device through which the network articulated what it actually wanted from data exchange; the ten consensus statements of 2024 are the crystallisation of that patient-first process. The requirements were thus articulated by those who carry the ultimate risk of any data architecture's failure, beginning from their independently formulated ideal, not derived from a technical proposal and presented to them for endorsement. The MPNE community intended Gillyweed as a design tool to capture and voice patient preferences, being fully aware that current platform models do not allow for its realisation. 
The Newtonian approach requires a custodian to hold the patient's identity state --- a rich state spanning the patient's doctors, health professionals, health records, family members, carers, and treatment history. For Gillyweed this would surface, before the patient had even begun, as the familiar onboarding ritual: \emph{create your account}, \emph{agree to our privacy policy and to how we use your data}. From the first step, the patient would know that an intermediary had become the custodian of their trust relations and that control over the use of their data had been delegated to that intermediary. Set against the Gillyweed vision, the gap is too wide to reconcile: the architecture contradicts the requirement at the moment of onboarding already.

However, the Relativistic approach proposed here actually allows regimes for considering MPNE's perspective stated in their statement 4 and 6. The Gillyweed vision of 2022 asks for an architecture in which the patient is the locus --- the reference frame --- of their own digital identity state. The patient is their own ecosystem, participating in and interacting with others, while remaining in control of what each relationship may do and of what they have consented to at any given moment. Where the custodial approach fails most visibly is in the propagation-lag regime of \S\ref{sec:fails}: maintaining consistency between how data is used and the consent that was given. In the relational approach, the primitives of \S\ref{sec:primitives} provide an infrastructure in which the patient's identity state can be reached by the respective systems of different health practitioners, each under the governance specific to that relationship --- and in which data flows not only between the patient and those systems but across practitioners, as patient-related data, without a custodian standing in the middle.

The four primitives map onto the Gillyweed requirements directly. The OOBIs that establish relations are as varied as the patient's ecosystem requires. DKMS resolves the keys that establish \emph{who} each party is, so that a clinician knows whether an exchange is with the patient directly or is brokered through a guardian. OCA supplies the semantics that fix \emph{what} a given exchange means. And distributed governance defines \emph{how} each relationship operates --- its modus operandi, its permissions, its constraints. In this composition, Gillyweed becomes something like a patient's data operating system: a personal control tower through which the patient directs the traffic of their own health data.
%
The relational approach is the natural soil for statements~4 and~6, which are architectural in character. The other consensus statements are not: they call for protections that technology alone cannot supply --- ethical, legal, and regulatory guardrails that belong to the community and its institutions. The architecture is built for exactly this division. It provides the structure within which such protections can be defined and enforced, without pretending to substitute for them. This is the architecture's contribution to statement~8, the community's insistence that technology cannot self-regulate and that control must also come through law: the architecture does not self-regulate either, but it makes the governance that regulates it explicit, auditable, and constitutive of every relation.

The MPNE engagement is significant beyond any technical artifact. The consensus was reached by a community that carries the ultimate risk of a data architecture's failure. It deliberated with access to multiple independent sources of expertise and reached its own conclusions, articulating requirements whose architectural core the custodial model cannot meet --- not because custodial systems have been built poorly, but because the regime in which patients now find themselves has moved beyond the assumption on which custodial systems rest. This is the root cause of the difficulty the community met when it tried to operationalise its vision as Gillyweed. The relational architecture answers those architectural requirements, while leaving room for the legal enforcement the community also demands. That a community deliberating independently should arrive at requirements the relational model satisfies and the custodial model structurally cannot is the strongest evidence we can offer that the model is responsive to a real and unmet need.

The lesson reaches beyond the melanoma community to the construction of the EHDS itself. As a domain-specific data space, the EHDS could be the demonstration that regulation enables data flows rather than blocking them --- a vibrant health economy operating at European scale, with all actors included. But the same dilemma returns: regulation fails to enable when the architecture beneath it does not match the problem the regulation was written to solve. The EHDS is being implemented now, at a pivotal moment in which the technical choices will determine whether the result protects patients and their communities or becomes one more layer of protection that serves neither the patient nor the health economy it is meant to support. It is a debate about implementation, not about the concept; and in it, the work of MPNE offers valuable insight.

\subsection{The Polish \emph{Truth on the Web} initiative: portable authenticity for public information}
\label{sec:truthontheweb}

Where the Swiss case treats the simultaneity of plural authority within a single sovereign act, the Polish \emph{Truth on the Web} initiative treats the propagation of authenticated public information beyond the custodian's domain --- an instance of the propagation-lag regime of \S\ref{sec:fails}, recast in the public sphere. The case is a finalist project in HackNation 2025, the implementation-driven hackathon organised by the Polish Ministry of Digital Affairs (Ministerstwo Cyfryzacji), under the patronage of Centralny O\'srodek Informatyki (COI), the institution responsible for Poland's deployed national digital-services and identity platform mObywatel. From 1{,}500 participants and over 430 submitted projects, the team Sigmion --- a collaboration between ArgonAuths and the Human Colossus Foundation, of which the authors are members --- was selected among the three finalists of the \emph{Prawda w Sieci} (\emph{Truth on the Web}) track, with an institutional evaluation path toward production deployment within COI \cite{argonauths2025hcf}.

The structural problem the initiative addresses is the following. Current architectures for trust in public information bind authentication to the \emph{location} from which information originates --- a government website, a recognised domain, a TLS certificate. This is a custodial architecture: the custodian's domain is the privileged frame from which authenticity can be read. The architecture works while the information remains within the custodian's domain, but fails the moment the information propagates outward: a screenshot of a government page can be altered before circulation, a PDF copied from an official source can be edited before re-distribution, a quotation can be paraphrased in transit, and AI-generated content can produce convincing forgeries of original material. The custodian's authority does not extend to verification at moments unconnected to the act of publication. This is the propagation-lag regime in its public-information form: the custodian's now is fixed to the moment of issuance, while the content's effective now continues through every subsequent presentation context.

The relational architecture proposed by Team Sigmion addresses this directly by binding authenticity to the \emph{information} rather than to its source location. Each item of public information carries a cryptographically grounded identifier and verifiable payload, such that any party --- citizen, journalist, archivist, downstream system --- can verify provenance independently of where or when the content is encountered. A screenshot taken from a government portal carries its verifiability with it; a PDF copied from an official source remains verifiable when redistributed; content quoted in another medium remains traceable to its issuing authority. The architecture composes naturally with mObywatel and is designed to integrate with the European Digital Identity Wallet framework under eIDAS 2.0, extending Poland's existing digital infrastructure from authenticated identity into authenticated public information.

The structural payoff of \S\ref{sec:relation} is again concretely visible. The Polish state retains its role as the authoritative issuer of public information, and the architecture preserves the strength of that issuance. But the architecture does not require the state to remain present at the moment of verification, because the verifiability has been moved from the custodian's domain into the intrinsic properties of the information itself. This is the intrinsic/extrinsic decomposition of \S\ref{sec:relation} applied to the public sphere: what the custodian issued is the intrinsic content (cryptographically grounded, end-verifiable), while the context of verification is extrinsic and may be supplied by any party in possession of the verifying material. The Polish case shows that the same structural primitive that handles citizen-state identity in the relational architecture also handles citizen-state public-information authenticity, and that the architecture scales from individual credentials to the public-sphere infrastructure of a national digital state.
\section{Conclusion}
\label{sec:conclusion}

Relativity did not abolish state. It redefined state as observer-relative, and in doing so unified phenomena that the Newtonian picture had been forced to treat as separate --- the propagation of light, the behaviour of fast-moving particles, the relationship between mass and energy. What had appeared as a list of distinct anomalies in the Newtonian regime turned out to be a single structural fact in the relativistic one.

The argument of this paper is that identity is in the same position. Custodial IAM has, with increasing effort, addressed a series of problems that it treats as separate pathologies: guardianship as an edge case to be patched onto the issuer-holder-verifier model; cross-border data flows as a question of federation protocols; consent management as a layer of dashboards above an unchanged substrate; e-government participation as a problem of platform security. The treatment compounds when the same domain admits structurally asymmetric framings: medical-data infrastructure under the conventional model places the researcher as the agent who accesses data and the patient as the subject whose data is governed (\S\ref{sec:nextgen}), or alternatively places the patient as the principal articulating their own requirements and the researcher as the entity who must request the data (\S\ref{sec:mpne}). Custodial IAM treats these as different problems requiring different solutions, because the power distribution across the custodial relations is different in each. The architecture we have described treats them as the same problem with the same answer: each is an autonomous principal holding its own frame; the relation between them is constituted explicitly under the ecosystem's governance; the asymmetry dissolves because no principal is structurally dependent on the other. The applications of \S\ref{sec:applications} are presented as evidence that this unification is operationally achievable, not as future scenarios but as present deployments and articulated requirements, demonstrated across cases that conventional framings would not have grouped together.

Distributed governance, on this account, does not abolish identity. It relocates the authority over identity to where it structurally belongs --- the autonomous principal --- and in doing so makes coherent the set of problems that custodial IAM has been treating one at a time. The regime in which this matters is not coming. It has arrived. The patient communities have articulated their requirements; the European research infrastructure is being built; the constitutional questions of digital signature collection are being posed inside federal institutions; the integrity of public information is being defended at national scale; and the autonomous agent literature is converging from another direction on the same primitive without yet recognising it. What the paper has offered is the structural account that ties these together and the architecture that operationalises it.

The argument so far has addressed how the principal becomes the locus of authority over its own identity. Two threads of further work follow directly from that result. The first concerns how relations between such principals are themselves governed --- the rules of interaction that constitute ecosystems as units of governance, the protocols by which those rules are themselves the object of agreement among the principals they constitute. This is the territory of a forthcoming Part~3 in which the governance of inter-principal protocols is treated directly. Part~1 established the autonomous principal; Part~2 has established that the principal must hold the reference frame; Part~3 will address the governance under which relations between such reference frames are constituted and evolve.

The second thread concerns the agentic regime named in \S\ref{sec:fails}. An autonomous agent acting on behalf of a principal --- and increasingly, on behalf of another agent --- presents the relational architecture with a case that is structurally the same as guardianship in reverse, but that opens its own programme: the verifiable provenance of delegation across chains of agents, the conditions under which an agent's authority is constituted and revoked, the governance of agent instantiation when the principal is itself constituted only for the duration of a task. The 2024--2026 literature on agent governance has been reaching for exactly the primitives that the relational architecture supplies; the work of making this connection rigorous is the subject of a parallel programme to Part~3, treated in dedicated forthcoming work.

%
\section*{Acknowledgements}
\label{sec:acknowledgements}

The authors are grateful to the institutional partners, individual contributors, and funding sources without whom the structural argument of this paper could not have been developed against the operational backdrop it describes. Responsibility for the argument and its presentation rests with the authors; the contributions named below are recognised for what they specifically enabled.

We thank the contributors at Human Colossus Foundation, whose mixed domain expertise shaped the work on the Distributed Governance Model, Overlays Capture Architecture, and decentralised key management infrastructure provides the technical substrate from which Part~2 is written. 

The cases in \S\ref{sec:applications} would not have been written without the institutional partners who have undertaken the practical work of building relational-identity systems. For the personalised medicine and patient-data infrastructure cases of \S\ref{sec:nextgen} and \S\ref{sec:mpne}, we acknowledge Bettina Ryll, Gilliosa Spurrier and the MPNE community and the participants of the NextGen project, too numerous to be cited individually. For the Swiss E-Collecting case of \S\ref{sec:ecollecting}, we acknowledge the Team~8 collaborators of the Federal Chancellery E-Collecting Hackathon 2025, including Damian Viz\'ar from CSEM, J\'erôme Campese from VOX Communication SA and Alessua Pacino for their participation; the structural reflections presented are the authors' alone and do not represent the positions of these collaborators or their institutions. For the Polish \emph{Truth on the Web} project of \S\ref{sec:truthontheweb}, we acknowledge the ArgonAuths team and the institutional context provided by Centralny O\'srodek Informatyki and the Polish Ministry of Digital Affairs (Ministerstwo Cyfryzacji).

We thank Ania Mitan (Human Colossus Foundation), Emrys Shoemaker (Global Governance Center - Geneva Graduate Institute), and Meri Valtiala (Human Colossus Foundation), whose readings of earlier drafts improved the precision of the argument at specific points.

\paragraph{Funding.} This work was supported by the NextGen project, which received funding from the European Union's Horizon Europe research and innovation programme under grant agreement No.~101136962. Views and opinions expressed are those of the authors only and do not necessarily reflect those of the European Union or the European Health and Digital Executive Agency (HADEA). Neither the European Union nor the granting authority can be held responsible for them. The Swiss participation in NextGen was funded by the Swiss Confederation through the State Secretariat for Education, Research and Innovation (SERI) under contract number~23.00540.

\bibliographystyle{unsrturl}
\bibliography{relativistic_iam}

\begin{thebibliography}{10}

\bibitem{page2023distributed}
Philippe Page, Paul Knowles, and Robert Mitwicki.
\newblock Distributed governance: a principal-agent approach to data governance
  --- {Part 1}: Background and core definitions.
\newblock arXiv:2308.07280, 2023.
\newblock URL: \url{https://arxiv.org/abs/2308.07280}.

\bibitem{w3c-did}
{World Wide Web Consortium (W3C)}.
\newblock Decentralized identifiers ({DIDs}) v1.0: Core architecture, data
  model, and representations.
\newblock Technical report, W3C Recommendation, 2022.
\newblock URL: \url{https://www.w3.org/TR/did-core/}.

\bibitem{w3c-vc}
{World Wide Web Consortium (W3C)}.
\newblock Verifiable credentials data model v2.0.
\newblock Technical report, W3C Recommendation, 2025.
\newblock URL: \url{https://www.w3.org/TR/vc-data-model-2.0/}.

\bibitem{allen2016path}
Christopher Allen.
\newblock The path to self-sovereign identity, 2016.
\newblock URL:
  \url{http://www.lifewithalacrity.com/2016/04/the-path-to-self-sovereign-identity.html}.

\bibitem{lamport1978time}
Leslie Lamport.
\newblock Time, clocks, and the ordering of events in a distributed system.
\newblock {\em Communications of the ACM}, 21(7):558--565, 1978.
\newblock \href {https://doi.org/10.1145/359545.359563}
  {\path{doi:10.1145/359545.359563}}.

\bibitem{shapiro2011crdts}
Marc Shapiro, Nuno Pregui\c{c}a, Carlos Baquero, and Marek Zawirski.
\newblock Conflict-free replicated data types.
\newblock In {\em Stabilization, Safety, and Security of Distributed Systems
  (SSS 2011)}, volume 6976 of {\em Lecture Notes in Computer Science}, pages
  386--400. Springer, 2011.
\newblock \href {https://doi.org/10.1007/978-3-642-24550-3_29}
  {\path{doi:10.1007/978-3-642-24550-3_29}}.

\bibitem{lamport1998paxos}
Leslie Lamport.
\newblock The part-time parliament.
\newblock {\em ACM Transactions on Computer Systems}, 16(2):133--169, 1998.
\newblock \href {https://doi.org/10.1145/279227.279229}
  {\path{doi:10.1145/279227.279229}}.

\bibitem{dkms-hcf}
{Human Colossus Foundation}.
\newblock {DKMS}: Decentralised key management system.
\newblock Project documentation.
\newblock URL: \url{https://dkms.colossi.network}.

\bibitem{smith2019keri}
Samuel~M. Smith.
\newblock Key event receipt infrastructure ({KERI}).
\newblock arXiv:1907.02143, 2021.
\newblock Version 2.59; original draft 2019.
\newblock URL: \url{https://arxiv.org/abs/1907.02143}.

\bibitem{toip-keri-spec}
{Trust over IP Foundation}.
\newblock Key event receipt infrastructure ({KERI}) specification.
\newblock Technical report, Trust over IP Foundation, KERI Specification
  Working Group, 2025.
\newblock Work in progress, kswg-keri-specification.
\newblock URL: \url{https://trustoverip.github.io/kswg-keri-specification/}.

\bibitem{mazieres2000sfs}
David Mazi\`eres.
\newblock {\em Self-certifying File System}.
\newblock PhD thesis, Massachusetts Institute of Technology, May 2000.
\newblock Introduces self-certifying pathnames, in which the identifier carries
  its own public key, removing the need for external key management.
\newblock URL: \url{https://pdos.csail.mit.edu/papers/mazieres-phd.ps.gz}.

\bibitem{oca-hcf}
{Human Colossus Foundation}.
\newblock {OCA}: Overlays capture architecture.
\newblock Specification and project documentation.
\newblock URL: \url{https://oca.colossi.network}.

\bibitem{smith-oobi-spec}
Samuel~M. Smith.
\newblock Out-of-band-introduction ({OOBI}) protocol.
\newblock IETF Internet-Draft draft-ssmith-oobi, 2023.
\newblock URL: \url{https://datatracker.ietf.org/doc/draft-ssmith-oobi/}.

\bibitem{foxit2012blacktulip}
{Fox-IT}.
\newblock Black tulip: Report of the investigation into the {DigiNotar}
  certificate authority breach.
\newblock Technical report, Fox-IT BV, August 2012.
\newblock Commissioned by the Dutch government following the 2011 DigiNotar
  compromise.
\newblock URL: \url{https://www.researchgate.net/publication/269333601}.

\bibitem{cfr-diginotar}
{Council on Foreign Relations}.
\newblock Compromise of certificate issuer {DigiNotar}.
\newblock CFR Cyber Operations Tracker, 2011.
\newblock URL:
  \url{https://www.cfr.org/cyber-operations/compromise-of-certificate-issuer-diginotar}.

\bibitem{comodo2011}
{Comodo Group}.
\newblock Report of incident --- comodo detected and thwarted an intrusion on
  26-mar-2011, March 2011.
\newblock URL:
  \url{https://www.comodo.com/Comodo-Fraud-Incident-2011-03-23.html}.

\bibitem{sovrin2019guardianship}
{Sovrin Foundation Guardianship Task Force}.
\newblock On guardianship in self-sovereign identity.
\newblock White paper, Sovrin Foundation, December 2019.
\newblock URL: \url{https://sovrin.org/library/guardianship-white-paper/}.

\bibitem{sovrin2021guardianship-impl}
{Sovrin Foundation Guardianship Working Group}.
\newblock Guardianship credentials implementation guidelines and technical
  requirements for guardianship in self-sovereign identity.
\newblock Technical report, Sovrin Foundation, April 2021.
\newblock URL:
  \url{https://sovrin.org/a-deeper-understanding-of-implementing-guardianship/}.

\bibitem{sovrin2023guardianship-v2}
{Sovrin Foundation Guardianship Working Group}.
\newblock On guardianship in self-sovereign identity, {V2}.
\newblock White paper, Sovrin Foundation, May 2023.
\newblock URL:
  \url{https://sovrin.org/release-update-guardianship-whitepaper-on-guardianship-in-self-sovereign-identity-v2/}.

\bibitem{dif2024guardianship}
{Decentralized Identity Foundation}.
\newblock Working group output on guardianship and delegation.
\newblock Decentralized Identity Foundation, 2024.
\newblock URL: \url{https://identity.foundation/}.

\bibitem{sovrin2024dissolution}
{Sovrin Foundation}.
\newblock The {Sovrin Foundation} has been dissolved but {Sovrin MainNet}
  remains.
\newblock Sovrin Foundation, 2024.
\newblock URL:
  \url{https://sovrin.org/the-sovrin-foundation-has-been-dissolved-but-sovrin-mainnet-remains/}.

\bibitem{icrc2022rfl}
{International Committee of the Red Cross}.
\newblock Sophisticated cyber-attack targets {Red Cross Red Crescent} data on
  500,000 people.
\newblock ICRC News Release, January 2022.
\newblock URL:
  \url{https://www.icrc.org/en/document/sophisticated-cyber-attack-targets-red-cross-red-crescent-data-500000-people}.

\bibitem{chaffer2024decentralized}
Tracey~Jaffe Chaffer, Charlton von Goins~II, Dylan Cotlage, Babatunde Okusanya,
  and Justin Goldston.
\newblock Decentralized governance of autonomous {AI} agents.
\newblock arXiv:2412.17114, 2024.
\newblock URL: \url{https://arxiv.org/abs/2412.17114}.

\bibitem{ruan2026agentcity}
Anbang Ruan and Xing Zhang.
\newblock {AgentCity}: Constitutional governance for autonomous agent economies
  via separation of power.
\newblock arXiv:2604.07007, 2026.
\newblock NetX Foundation.
\newblock URL: \url{https://arxiv.org/abs/2604.07007}.

\bibitem{nextgen-cordis}
{NextGen Consortium}.
\newblock {NextGen}: Secured data integration in cardiovascular personalised
  medicine.
\newblock EU Horizon Europe project, grant agreement ID 101136962.
\newblock URL: \url{https://cordis.europa.eu/project/id/101136962}.

\bibitem{chahal2025interop}
C.~Anwar~A. Chahal, Fares Alahdab, Babken Asatryan, Daniel Addison, Nay Aung,
  Mina~K. Chung, Spiros Denaxas, Jessilyn Dunn, Jennifer~L. Hall, Nathalie
  Pamir, David~J. Slotwiner, Jose~D. Vargas, and Antonis~A. Armoundas.
\newblock Data interoperability and harmonization in cardiovascular genomic and
  precision medicine.
\newblock {\em Circulation: Genomic and Precision Medicine}, 18(6):e004624,
  2025.
\newblock URL:
  \url{https://www.ahajournals.org/doi/10.1161/CIRCGEN.124.004624}, \href
  {https://doi.org/10.1161/CIRCGEN.124.004624}
  {\path{doi:10.1161/CIRCGEN.124.004624}}.

\bibitem{ehds-regulation}
{European Union}.
\newblock Regulation ({EU}) 2025/327 of the {European Parliament} and of the
  {Council} of 11 {February} 2025 on the {European Health Data Space}.
\newblock Official Journal of the European Union, 2025.
\newblock URL: \url{https://eur-lex.europa.eu/eli/reg/2025/327/oj}.

\bibitem{schlink1986volkszaehlungsurteil}
Bernhard Schlink.
\newblock Das {Volksz{\"a}hlungsurteil} des {Bundesverfassungsgerichts} und die
  {Frage} nach dem {Recht} auf informationelle {Selbstbestimmung}.
\newblock In {\em Der Staat}, volume~25, pages 233--252. -, 1986.
\newblock Reference work on the German Federal Constitutional Court's 1983
  Census Judgement establishing the right to informational self-determination.

\bibitem{rouvroy2009right}
Antoinette Rouvroy and Yves Poullet.
\newblock The right to informational self-determination and the value of
  self-development: Reassessing the importance of privacy for democracy.
\newblock In Serge Gutwirth, Yves Poullet, Paul De~Hert, C{\'e}cile
  de~Terwangne, and Sjaak Nouwt, editors, {\em Reinventing Data Protection?},
  pages 45--76. Springer, 2009.
\newblock URL:
  \url{https://link.springer.com/chapter/10.1007/978-1-4020-9498-9_2}, \href
  {https://doi.org/10.1007/978-1-4020-9498-9_2}
  {\path{doi:10.1007/978-1-4020-9498-9_2}}.

\bibitem{nextgen-mmio}
{NextGen Consortium}.
\newblock Multimodal integration objects ({MMIOs}).
\newblock NextGen project tools documentation, 2026.
\newblock URL:
  \url{https://www.nextgentools.eu/tools/multimodal-integration-objects-mmios/}.

\bibitem{wilkinson2016fair}
Mark~D. Wilkinson, Michel Dumontier, IJsbrand~Jan Aalbersberg, Gabrielle
  Appleton, Myles Axton, Arie Baak, Niklas Blomberg, Jan-Willem Boiten,
  Luiz~Bonino da~Silva~Santos, Philip~E. Bourne, et~al.
\newblock The {FAIR} {Guiding} {Principles} for scientific data management and
  stewardship.
\newblock {\em Scientific Data}, 3:160018, 2016.
\newblock \href {https://doi.org/10.1038/sdata.2016.18}
  {\path{doi:10.1038/sdata.2016.18}}.

\bibitem{swiss-constitution-139}
{Swiss Confederation}.
\newblock Federal constitution of the swiss confederation, article 139: Popular
  initiative for a partial revision of the federal constitution, 1999.
\newblock SR 101, as amended.
\newblock URL: \url{https://www.fedlex.admin.ch/eli/cc/1999/404/fr}.

\bibitem{rts2024fraud}
{Radio T\'el\'evision Suisse}.
\newblock Enqu\^ete sur une possible fraude lors de la collecte de signatures
  pour des initiatives populaires.
\newblock RTS Info, September 2024.
\newblock URL:
  \url{https://www.rts.ch/info/suisse/2024/article/enquete-sur-une-possible-fraude-lors-de-la-collecte-de-signatures-pour-des-initiatives-populaires-28617919.html}.

\bibitem{cf2024ecollecting}
{Conseil F\'ed\'eral Suisse}.
\newblock R\'ecolte \'electronique des signatures \`a l'appui des initiatives
  populaires et des demandes de r\'ef\'erendum au niveau f\'ed\'eral.
\newblock Rapport du conseil f\'ed\'eral, Conseil F\'ed\'eral Suisse, Berne,
  November 2024.
\newblock URL:
  \url{https://www.newsd.admin.ch/newsd/message/attachments/90668.pdf}.

\bibitem{swiss-ecollecting-2025}
{Swiss Federal Chancellery}.
\newblock {E-Collecting Hackathon 2025}, {Team 8}: Instaurer la confiance \`a
  chaque signature --- collecte \'electronique f\'ed\'er\'ee et v\'erifi\'ee
  cryptographiquement, October 2025.
\newblock Contributors: P. Page, R. Mitwicki, J. Campese, A. Pacino, D.
  Viz\'ar, M. Pietrus.
\newblock URL: \url{https://github.com/swiss/e-collecting-hackathon-team8}.

\bibitem{page2025evoting}
Philippe Page.
\newblock Switzerland's {E-Challenges} --- and what the world can learn.
\newblock Human Colossus Foundation, November 2025.
\newblock URL:
  \url{https://humancolossus.foundation/blog/voting-without-tracing}.

\bibitem{itobos-cordis}
{iToBoS Consortium}.
\newblock {iToBoS}: Intelligent total body scanner for early detection of
  melanoma.
\newblock EU Horizon 2020 project, grant agreement ID 965221.
\newblock URL: \url{https://cordis.europa.eu/project/id/965221}.

\bibitem{mpne2024consensus}
{Melanoma Patient Network Europe}.
\newblock Patient consensus on data, {AI} and data-dependent models for
  business and research.
\newblock White paper, Melanoma Patient Network Europe (MPNE), 2024.
\newblock MPNEconsensus 2024, Berlin.
\newblock URL:
  \url{https://www.mpneurope.org/patient-consensensus-data-and-ai}.

\bibitem{spurrier2022sync}
Gilly Spurrier.
\newblock Sync not sink our data: We need more cancer patient agency in health
  data use.
\newblock Melanoma Patient Network Europe blog, October 2022.
\newblock URL:
  \url{https://www.mpneurope.org/post/sync-not-sink-our-data-we-need-more-cancer-patient-agency-in-health-data-use}.

\bibitem{mpne-gillyweed}
{Melanoma Patient Network Europe}.
\newblock {GILLYWEED}: What patients want from their health data.
\newblock MPNEminiMeet2022 workshop materials, Brussels, November 2022.
\newblock Patient-authored design fiction for the European Health Data Space;
  basis for the MPNEconsensus 2024 process.
\newblock URL: \url{https://www.mpneurope.org/mpne2022}.

\bibitem{argonauths2025hcf}
Philippe Page, {ArgonAuths}, and {Human Colossus Foundation}.
\newblock {ArgonAuths} x {Human Colossus}: Finalists at {HackNation} 2025 ---
  redefining trust in the digital public sphere.
\newblock Human Colossus Foundation Blog, December 2025.
\newblock URL:
  \url{https://humancolossus.foundation/blog/congrats-to-our-team-}.

\end{thebibliography}

\end{document}